\documentclass[
reprint,
superscriptaddress,
 amsmath,amssymb,
 aps,
prl,
]{revtex4-2}
\usepackage{graphicx}
\usepackage{dcolumn}
\usepackage{bm}
\usepackage{physics}
\usepackage{hyperref}
\usepackage{amsmath}
\hypersetup{
     colorlinks   = true,
     linkcolor    = blue,
     citecolor    = blue,
	 urlcolor     = blue
}

\begin{document}

\title{First-principles electron-phonon interactions and electronic transport\\
in large-angle twisted bilayer graphene}

\author{Shiyuan Gao}
\affiliation{Department of Applied Physics and Materials Science, and Department of Physics,\\ California Institute of Technology, Pasadena, California 91125, USA}
\affiliation{Department of Physics and Astronomy, Institute for Quantum Matter,\\ Johns Hopkins University, Baltimore, Maryland 21218, USA}

\author{Jin-Jian Zhou}
\affiliation{School of Physics, Beijing Institute of Technology, Beijing 100081, China}

\author{Yao Luo}
\affiliation{Department of Applied Physics and Materials Science, and Department of Physics,\\ California Institute of Technology, Pasadena, California 91125, USA}

\author{Marco Bernardi}
\email{bmarco@caltech.edu}
\affiliation{Department of Applied Physics and Materials Science, and Department of Physics,\\ California Institute of Technology, Pasadena, California 91125, USA}

\begin{abstract} 
Twisted bilayer graphene (tBLG) has emerged as an exciting platform for novel \mbox{condensed matter} physics. However, electron-phonon ($e$-ph) interactions in tBLG and their effects on electronic transport are not completely understood. Here we show first-principles calculations of 
$e$-ph interactions and resistivity in commensurate tBLG with large twist angles of 13.2 and 21.8 degrees. 
These calculations overcome key technical barriers, including large unit cells of up to 76 atoms, Brillouin-zone folding of the $e$-ph interactions, and unstable lattice vibrations due to the AA-stacked domains. 
We show that $e$-ph interactions due to layer-breathing (LB) phonons enhance intervalley scattering in large-angle tBLG. This interaction effectively couples the two layers, which are otherwise electronically decoupled at such large twist angles. 
As a result, the phonon-limited resistivity in tBLG deviates from the temperature-linear trend characteristic of monolayer graphene and tBLG near the magic angle. 
Taken together, our work quantifies $e$-ph interactions and scattering mechanisms in tBLG, revealing subtle interlayer coupling effects at large twist angles. 
\end{abstract}
\maketitle

\vspace{-10pt}
Electronic transport in graphene is a rich subject \cite{DasSarma2011}. Theory showed that the carrier mobility in graphene is limited by acoustic phonons~\cite{Hwang2008}, and then focused on bilayer graphene \cite{Min2011} and the role of \mbox{flexural phonons~\cite{Mariani2008,Mariani2010,Castro2010}.} 
First-principles calculations based on density functional theory (DFT) have also been employed to study phonon-limited transport in graphene, initially using the deformation potential approximation \cite{Borysenko2010,Borysenko2011,Li2011,Kaasbjerg2012} and later with first-principles $e$-ph interactions~\cite{Park2014,Sohier2014}, which take into account all phonon modes and electronic states on equal footing, enabling quantitative \mbox{predictions.}
\\
\indent
Twisted bilayer graphene (tBLG) is now attracting intense interest due to the emergence of correlated insulating and superconducting states near the magic \mbox{angle~\cite{Cao2018,Cao2018a}.} 
While the origin of these phases is still debated, \mbox{$e$-ph} interactions are thought to play an important role in the rich physics of tBLG~\cite{Choi2018,Peltonen2018,Wu2018,Lian2019,Andrei2020}. 
For example, \mbox{electronic transport} is intimately linked to the $e$-ph interactions in tBLG, and experiments have observed a large linear-in-temperature resistivity near the magic angle~\cite{Polshyn2019}, whose origin remains unclear.
\\
\indent
Theoretical work has focused on analytic and tight-binding models of $e$-ph interactions and phonon-limited resistivity in tBLG \cite{Ray2016, Choi2018, Ochoa2019,Wu2019,Li2020}. The effects of lattice relaxation on the electronic structure of tBLG have also been studied using both tight-binding and continuum models~\cite{Koshino2017, Kaxiras2019, Wagner2022}. 
\mbox{However,} explicit first-principles calculations of $e$-ph interactions in tBLG have remained \mbox{prohibitive} due to the large unit cell sizes, particularly at small twist angles. 
\\
\indent
Here, we show first-principles calculations of $e$-ph interactions and transport properties in large-angle tBLG. Our approach allows us to quantify the contributions to electron scattering and transport from different acoustic and optical modes and from intravalley and intervalley processes. Similar to monolayer graphene, we find that the resistivity in large-angle tBLG is controlled by acoustic phonon scattering at low temperature and optical phonon scattering near and above room temperature. Yet, the phonon-limited resistivity in tBLG is found to deviate significantly from two decoupled layers, with a faster-than-linear temperature dependence due to intervalley scattering mediated by layer-breathing phonons. These results reveal subtle interactions emerging from the twist-angle degree of freedom, highlighting the promise of first-principles calculations to study tBLG.
\\
\indent
The commensurate moir\'e superlattice of tBLG is generated by rotating the superlattice vector of one graphene layer, $\mathbf{a}_S^{(1)}\!=\!m\mathbf{a}_1+n\mathbf{a}_2$, into $\mathbf{a}_S^{(2)}\!=\!n\mathbf{a}_1+m\mathbf{a}_2$, where $\mathbf{a}_1$ and $\mathbf{a}_2$ are graphene lattice vectors and $(m, n)$ is a pair of co-prime integers \cite{Mele2010,Moon2013}. We focus on the two smallest tBLG unit cells, a 28-atom unit cell with $(m,n)\!=\!(2,1)$ and a 76-atom unit cell with $(m, n)\!=\!(3, 2)$, which correspond to twist angles of 21.8$^{\circ}$ and 13.2$^{\circ}$ respectively.
We compute the ground state electronic structure of these tBLG systems using DFT in a plane-wave basis with the \textsc{Quantum ESPRESSO} package~\cite{giannozzi_quantum_2009,giannozzi_advanced_2017}. We employ the local density approximation with a plane-wave kinetic energy cutoff of 90 Ry, using the experimental lattice constant of 2.46 $\mathrm{\AA}$ for graphene. The resulting electronic band structures of 21.8$^{\circ}$ and 13.2$^{\circ}$ tBLG are given in the Supplemental Material (SM)~\cite{supp}.
\\
\indent
The key quantities in first-principles $e$-ph calculations are the $e$-ph coupling matrix elements \cite{Zhou2021}
\vspace{-4pt}
\begin{equation}
\label{eq:g_eph}
\!\!\!\!\!g_{mn\nu}(\mathbf{k,q})=\sqrt{\frac{\hbar}{2\omega_{\nu\mathbf{q}}}}
\sum_{\kappa\alpha}\frac{\mathbf{e}^{\kappa\alpha}_{\nu\mathbf{q}}}{\sqrt{M_\kappa}}
\mel{m\mathbf{k}+\mathbf{q}}{\partial_{\mathbf{q}\kappa\alpha}V}{n\mathbf{k}},
\end{equation}
which describe the probability amplitude for an electron to scatter from a band state $\ket{n\mathbf{k}}$ to state $\ket{m\mathbf{k}+\mathbf{q}}$ by emitting or absorbing a phonon with \mbox{momentum $\mathbf{q}$,} mode index $\nu$, energy $\hbar\omega_{\nu\mathbf{q}}$ and eigenvector $\mathbf{e}^{\kappa\alpha}_{\nu\mathbf{q}}$. The \mbox{perturbation potential} $\partial_{\mathbf{q}\kappa\alpha}V$ is the change in the DFT Kohn-Sham potential for a unit displacement of atom $\kappa$ (with mass $M_\kappa$) in the Cartesian direction $\alpha$~\cite{Zhou2021}, and is obtained from density functional perturbation theory (DFPT). 
We use the {\sc Perturbo} code \cite{Zhou2021} to compute $g_{mn\nu}(\mathbf{k,q})$ at $\mathbf{k}$- and $\mathbf{q}$-points on fine grids, starting from DFPT results on $6\times6\times1$ grids in 21.8$^{\circ}$ tBLG and $3\times3\times1$ grids in 13.2$^{\circ}$ tBLG for both $\mathbf{k}$- and $\mathbf{q}$-points. The interpolation of the $e$-ph matrix elements uses maximally localized Wannier functions obtained with the {\sc Wannier90} code \cite{Pizzi2020}. 
\begin{figure}[t]
\centering
\includegraphics[width=8.6cm,clip]{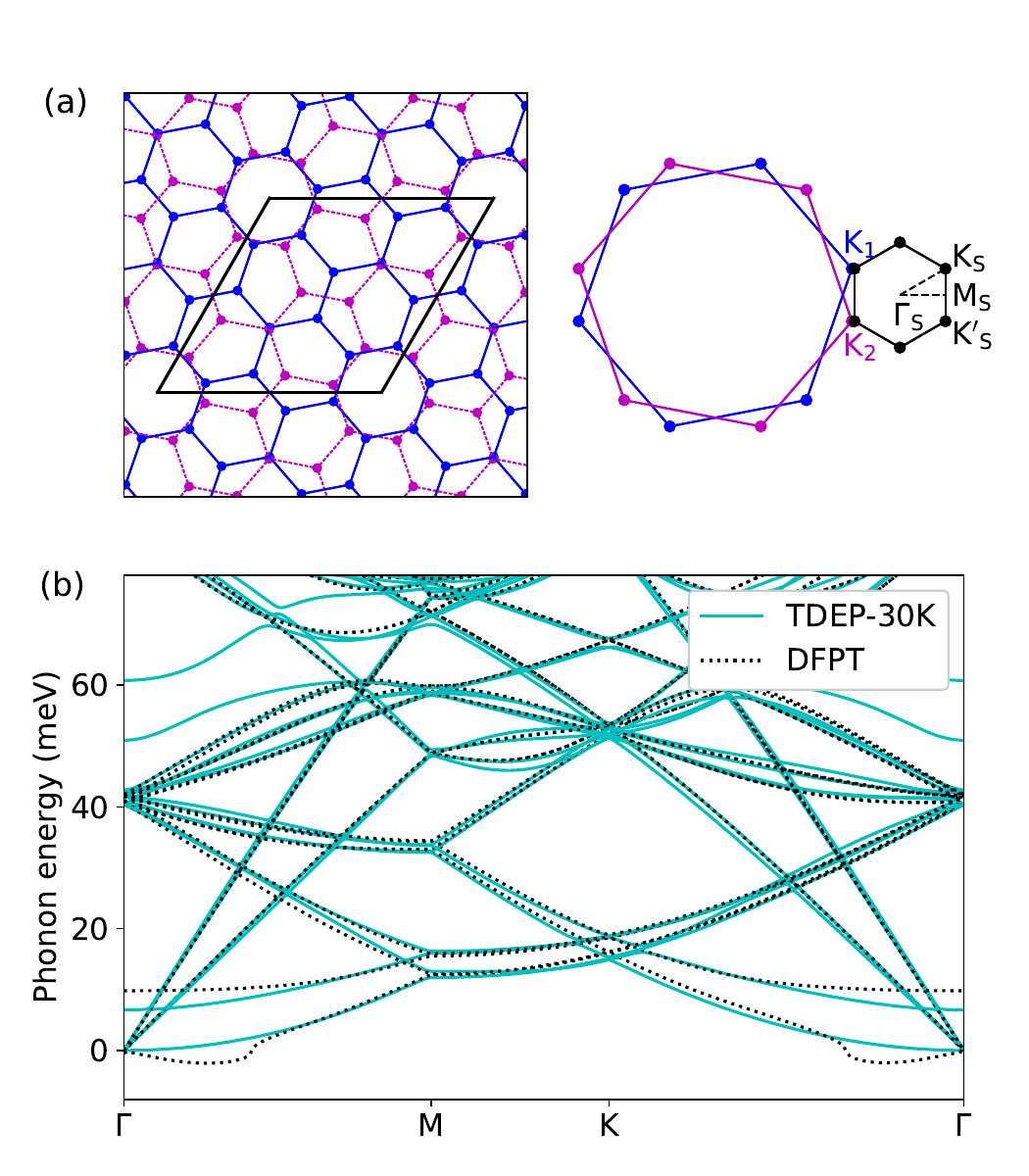}
\caption{(a) Illustration of the crystal structure and Brillouin zone of 21.8$^{\circ}$ tBLG.  (b) Low-energy phonon dispersions in 21.8$^{\circ}$ tBLG, calculated using TDEP at 30 K (solid line) and DFPT (dotted line). }
\label{fig:tdep}
\end{figure}

Figure \ref{fig:tdep}(a) shows the crystal structure and Brillouin zone of tBLG with 21.8$^{\circ}$ twist angle. Because the structure of tBLG is energetically unfavorable, the zero-temperature phonon dispersions computed with DFPT exhibit imaginary low-energy phonons at long-wavelength, shown as negative energies in the DFPT curves in Fig.~\ref{fig:tdep}(b). This is an evidence of dynamical instability of tBLG~\cite{Pallikara_2022}, where the imaginary phonons are soft modes associated with structural distortions. 
\\
\indent
To treat these soft modes, we use the temperature-dependent effective potential (TDEP) method \cite{Hellman2011,Hellman2013}, which provides effective harmonic inter-atomic force constants at a given temperature from a least-square fit of the DFT forces. 
These calculations use 252-atom supercells with atomic positions generated stochastically according to the canonical ensemble at each temperature. (As the phonons are determined by the force constants, this process is iterated until convergence~\cite{Hellman2011,Hellman2013}.)
The resulting phonon dispersions are free of imaginary modes, as shown in the TDEP curves at 30~K in Fig.~\ref{fig:tdep}(b) and at 300~K in the SM~\cite{supp}.
\begin{figure}[!t]
\centering
\includegraphics[width=8.6cm,trim={0cm 0cm 0cm 0cm},clip]{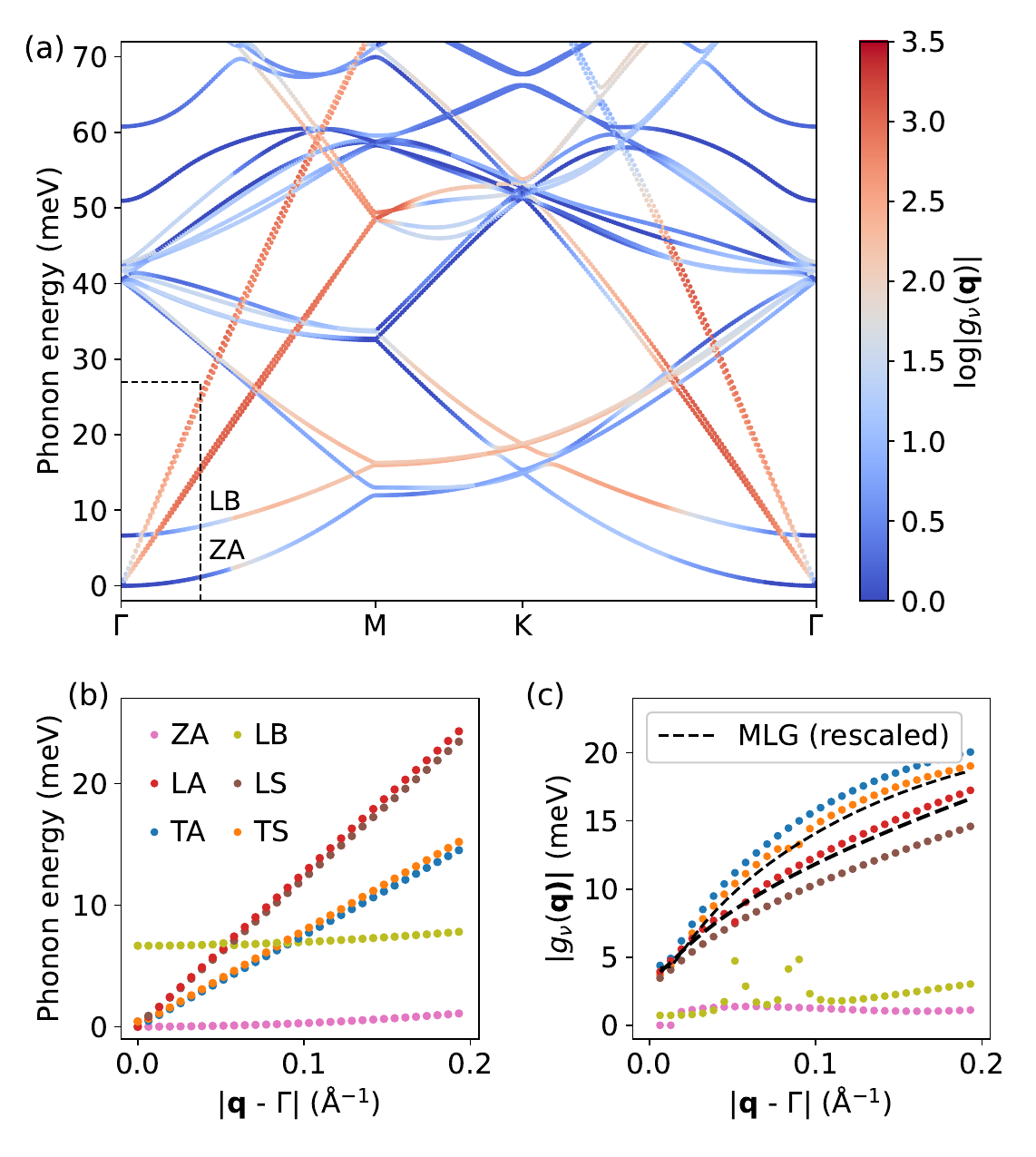}
\caption{(a) Phonon dispersions in 21.8$^{\circ}$ tBLG computed at 30 K using TDEP and color-coded according to the log of the $e$-ph coupling strength, $\log|g_\nu(\mathbf{q})|$. The dashed box is the region plotted in panel (b). (b) Low-energy phonon dispersions near the $\Gamma$ point, with different phonon modes shown in different colors. (c) The $e$-ph coupling strength $|g_\nu(\mathbf{q})|$ as a function of phonon wave-vector $\mathbf{q}$ with the same coloring scheme as in (b). The $e$-ph coupling strengths for the TA and LA modes in MLG, rescaled by an appropriate unit-cell size factor (see SM~\cite{supp}), are shown with dashed lines for comparison.}
\label{fig:gkq}
\end{figure}

Addressing the soft modes allows us to more precisely study the $e$-ph interactions with low-energy phonons~\cite{zhou_soft}. We recompute the $e$-ph matrix elements in Eq.~(\ref{eq:g_eph}) by replacing the phonon frequencies $\omega_{\nu\mathbf{q}}$ and eigenvectors $\mathbf{e}^{\kappa\alpha}_{\nu\mathbf{q}}$ with corresponding quantities obtained from TDEP. 
\mbox{Figure~\ref{fig:gkq}(a)} shows the phonon dispersions in 21.8$^{\circ}$ tBLG color-coded according to the $e$-ph coupling strength, $|g_\nu(\mathbf{q})|=(\sum_{mn}|g_{mn\nu}(\mathbf{k}=K,\mathbf{q})|^2/N_b)^{1/2}$, computed by averaging the contributions of the $N_b\!=\!4$ bands crossing the Dirac point. 
At low phonon energy, we find that the  acoustic phonons with linear dispersions exhibit the strongest e-ph coupling. The other two low-energy modes, with energy below 20 meV, are out-of-plane layer-breathing (LB) and flexural (ZA) phonons with quadratic dispersions at long wavelength~\cite{Ray2016}. The LB mode, which has a finite energy at the zone center, exhibits sizable $e$-ph coupling strengths, particularly for phonon wave-vectors near the K point of the Brillouin zone, while the ZA mode has a negligible $e$-ph coupling, similar to MLG. 
\\
\indent
We further analyze the $e$-ph coupling strengths for these different low-energy phonons. Each graphene layer contributes two acoustic and one flexural mode, and thus tBLG has a total of six low-energy modes. In Fig.~\ref{fig:gkq}(b)-(c), we label the low-energy modes according to their character, obtained by projecting the eigenvectors onto six distinct displacement patterns: longitudinal acoustic (LA) and transverse acoustic (TA) modes, where the two layers move in the same in-plane direction; longitudinal shearing (LS) and transverse shearing (TS) modes, where the two layers move in opposite in-plane directions; and out-of-plane modes, where the two layers move in the same (ZA mode) or opposite (LB mode) layer-normal directions. Below, for simplicity we refer to these six low-energy modes as the acoustic phonons, and the remaining modes as optical phonons, in tBLG.
\\
\indent
This classification allows us to compare the $e$-ph coupling strength in tBLG with monolayer graphene (MLG) [see the dashed lines in Fig.~\ref{fig:gkq}(c)], which also requires a proper rescaling of the MLG $e$-ph coupling (see SM~\cite{supp}). The LA and TS modes in tBLG have $e$-ph coupling strengths almost identical to the LA and TA modes in MLG, respectively. In addition, the LS and TA modes in tBLG have a slightly smaller and greater coupling strengths than the LA and TA modes in MLG, namely their respective MLG counterparts. 
Similar to MLG, the flexural ZA mode couples weakly with electrons. The coupling is stronger for the LB mode, but overall the $e$-ph coupling for the ZA and LB out-of-plane modes in tBLG is weaker than for the in-plane modes. These results show that $e$-ph interactions for acoustic phonons in large-angle tBLG are overall similar $-$ except for the LB mode $-$ to two decoupled \mbox{graphene layers.}
\begin{figure*}[t]
\centering
\includegraphics[width=17.2cm,trim={0cm 0cm 0cm 0cm},clip]{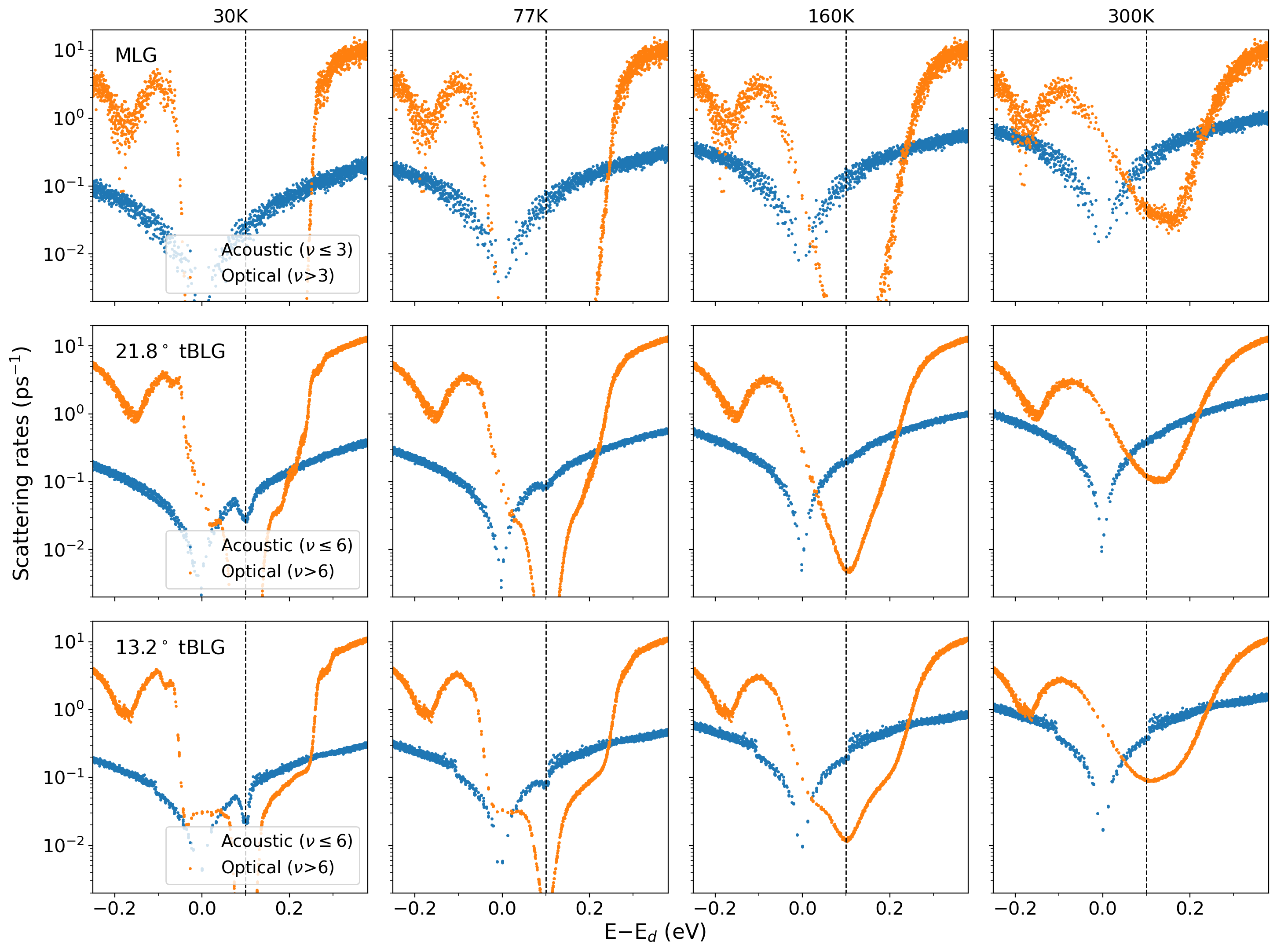}
\caption{Electron-phonon scattering rate as a function of electron energy, shown separately for the acoustic and optical modes, in MLG and tBLG with 21.8$^{\circ}$ and 13.2$^{\circ}$ twist angles. We compare results at four different temperatures between 30 $-$ 300 K. The energy zero is set to the Dirac point and the Fermi energy, shown with a vertical line, is 0.1 eV above the Dirac point.}
\label{fig:scatter rate}
\end{figure*}
\\
\indent
Starting from the $e$-ph matrix elements, we use Fermi's golden rule to compute the $e$-ph scattering rate for each electronic state~\cite{Zhou2021,bernardi2016}:
\begin{equation} \label{eq:scatrate}
\begin{split}
\Gamma_{n\mathbf{k}}(T) = &\frac{2\pi}{\hbar \mathcal{N}_\mathbf{q}}\sum_{m \nu \mathbf{q}}\abs{g_{mn\nu}(\mathbf{k},\mathbf{q})}^2 \\
& [(N_{\nu \mathbf{q}}+1-f_{m\mathbf{k}+\mathbf{q}})\delta(\varepsilon_{n\mathbf{k}}-\varepsilon_{m\mathbf{k}+\mathbf{q}}-\hbar\omega_{\nu \mathbf{q}}) \\
& + (N_{\nu \mathbf{q}}+f_{m\mathbf{k}+\mathbf{q}})\delta(\varepsilon_{n\mathbf{k}}-\varepsilon_{m\mathbf{k}+\mathbf{q}} +\hbar\omega_{\nu \mathbf{q}})],
\end{split}
\end{equation}
where $\mathcal{N}_\mathbf{q}$ is the number of $\mathbf{q}$-points, and $N_{\nu\mathbf{q}}$ and $f_{m\mathbf{k+q}}$ are, respectively, equilibrium phonon and electron occupations at temperature $T$~\footnote{We use $10^6$ random $\mathbf{q}$-points sampled uniformly in the Brillouin zone and a 2 meV Gaussian smearing for the Dirac delta function to ensure the scattering rate is converged down to 15~K.}. 
\\
\indent 
Figure~\ref{fig:scatter rate} shows the scattering rate as a function of electron energy for MLG and for tBLG with 21.8$^{\circ}$ and 13.2$^{\circ}$ twist angles. We separate the contributions to scattering from acoustic and optical phonons. 
Acoustic phonon scattering follows a similar behavior in MLG and large-angle tBLG: in both systems, the acoustic scattering rate vanishes at the Dirac point but is finite at the Fermi energy. The acoustic scattering rate decreases rapidly as the electron energy approaches the Dirac point due to the small electronic density of states, which is proportional to the phase space for acoustic $e$-ph scattering~\cite{Marco2014}. 
\\
\indent
Near the Fermi energy, in both MLG and large-angle tBLG, the acoustic $e$-ph scattering rate is significant, of order $\sim$$0.01 - 0.1$ ps$^{-1}$ in the 30$-$300~K temperature range studied here.
Conversely, both in MLG and large-angle tBLG, the scattering rate for optical phonons is exponentially suppressed at low temperature near the Fermi energy due to the greater thermal energy required to excite optical phonons. At higher temperatures (160 K and 300~K) the optical-phonon scattering rates are on average of order 0.1 ps$^{-1}$ near the Fermi energy, and thus comparable to the acoustic phonon scattering rates. 
Interestingly, these results show that carrier scattering with acoustic and optical phonons are governed by two distinct energies $-$ the Dirac point for acoustic phonons, and the Fermi energy for optical phonons. This implies that doping can tune their relative contributions to transport in large-angle tBLG.

\begin{figure*}[t]
\centering
\includegraphics[width=17.5cm,trim={0cm 0cm 0cm 0cm},clip]{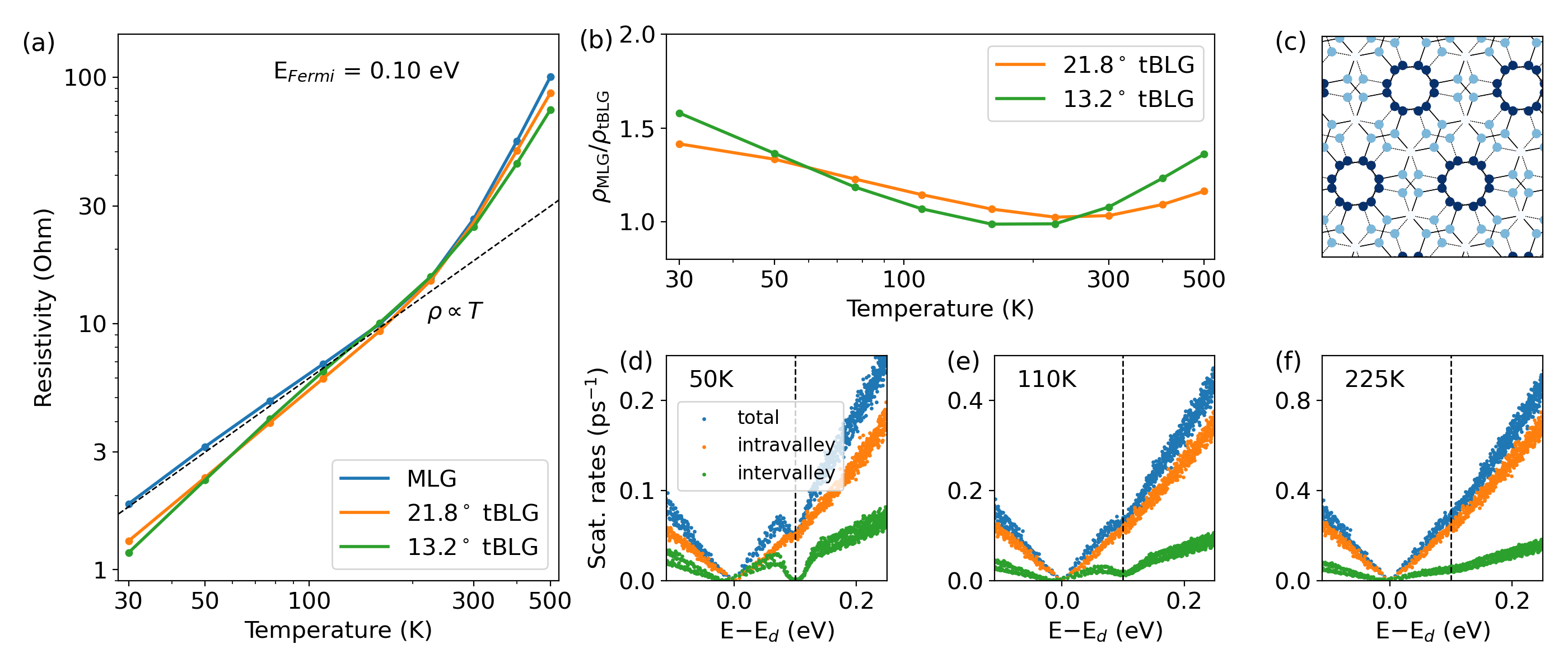}
\caption{(a) Phonon-limited resistivity versus temperature in MLG and large-angle tBLG with 21.8$^{\circ}$ and 13.2$^{\circ}$ twist angles. The Fermi level is set to 0.1 eV above the Dirac point. The black dashed line guides the eye and indicates a $T$-linear resistivity. (b) Resistivity ratio between MLG and tBLG as a function of temperature. (c) Atomic displacement pattern for the LB phonon mode governing intervalley scattering in 21.8$^{\circ}$ tBLG. The color of the atoms indicates the magnitude of the displacement (obtained by analyzing the phonon eigenvector) with darker blue colors corresponding to greater layer-normal displacements. The direction of the motion is layer-normal with atoms in different layers moving in opposite directions. (d)-(f) Electron-phonon scattering rates as a function of carrier energy for 21.8$^{\circ}$ tBLG at three temperatures, showing the contributions of acoustic phonons to intravalley scattering ($\nu\leq6, |\mathbf{q}|<\frac{2\pi}{3a}$) and intervalley scattering ($\nu\leq6, |\mathbf{q}|>\frac{2\pi}{3a}$) separately.}
\label{fig:trans}
\end{figure*}

To study electronic transport, we compute the phonon-limited conductivity using the Boltzmann transport equation in the relaxation time approximation~\cite{Zhou2021}: 
\begin{equation}
\centering
\sigma_{\alpha\beta}(T) = \frac{e^2S}{N_\mathbf{k}\Omega}\sum_{n\mathbf{k}}\tau_{n\mathbf{k}}(T)\, v_{n\mathbf{k}}^\alpha v_{n\mathbf{k}}^\beta\left(-\frac{\partial f_{n\mathbf{k}}(T)}{\partial\varepsilon_{n\mathbf{k}}}\right) ,
\end{equation}
where $\alpha$ and $\beta$ are Cartesian directions, $\Omega$ is the unit cell volume, $S$ is the spin degeneracy, $v_{n\mathbf{k}}$ are band velocities, $\tau_{n\mathbf{k}}=1/\Gamma_{n\mathbf{k}}$ are $e$-ph relaxation times (i.e., the inverses of the $e$-ph scattering rates) at temperature $T$, and $f_{n\mathbf{k}}$ are equilibrium Fermi-Dirac occupations. In these calculations, we include the four bands intersecting at the Dirac point, sampling an energy window within 0.35 eV of the Fermi level with a dense $\mathbf{k}$-point grid corresponding to 1500$\times$1500 $\mathbf{k}$-points in the Brillouin zone of MLG (the number of $\mathbf{k}$-points is rescaled according to the Brillouin zone size of tBLG).  
\\
\indent
Figure~\ref{fig:trans}(a) shows the computed in-plane resistivity, $\rho \equiv2/(\sigma_{xx}+\sigma_{yy})$, as a function of temperature for MLG and large-angle tBLG with 21.8$^{\circ}$ and 13.2$^{\circ}$ twist angles. The lowest temperature considered here (30 K) is close to the Bloch-Gr\"uneisen temperature $T_\mathrm{BG}$~\cite{Stormer1990}, where $k_\mathrm{B}T_\mathrm{BG}$ equals the energy of the acoustic phonon with wave-vector $2k_F$ ($T_\mathrm{BG}\approx 37$~K based on our calculations)~\footnote{At temperatures below $T_\mathrm{BG}$, which are not consider here, the resistivity approaches a $\rho\propto T^4$ behavior in the low-temperature limit~\cite{Hwang2008}.}.
Above $T_\mathrm{BG}$, there is a wide temperature window up to $\sim$250 K where the resistivity is limited by acoustic-phonon scattering and appears as a straight line in the log-log plot in Fig.~\ref{fig:trans}(a). 
\\
\indent
This trend breaks down above $\sim$250 K, where scattering from optical phonons becomes important and governs the resistivity, which converges to a similar value in MLG and the two large-angle tBLG studied here. At intermediate temperatures between 30$-$250~K, $\rho$ is linear in $T$ in MLG, in agreement with previous theoretical predictions~\cite{Hwang2008,Min2011} and experiments~\cite{Dean2010, Efetov2010}. 
In contrast, the resistivity in large-angle tBLG is found to deviate from a linear temperature trend. 
By fitting the data to a power law, we find $\rho \propto T^{1.17}$ for 21.8$^{\circ}$ tBLG and $\rho \propto T^{1.28}$ for 13.2$^{\circ}$ tBLG in this temperature range. This differs from the behavior of tBLG near the magic angle, where the experimental resistivity is $T$-linear at these \mbox{temperatures~\cite{Polshyn2019}.} 
\\
\indent
To better understand this non-linear temperature dependence in large-angle tBLG, in Fig.~\ref{fig:trans}(b) we plot the ratio of the MLG and tBLG resistivities, $\rho_{\rm MLG}/\rho_{\rm tBLG}$, versus temperature. The resulting curve is concave up, approaching a value of about 1.5 at low temperature and minimum values of 1.1 at $\sim$300 K for 21.8$^{\circ}$ tBLG and 1.0 at $\sim$200 K for 13.2$^{\circ}$ tBLG. If the two layers in tBLG were fully decoupled, from a parallel-resistor argument the resistivity ratio $\rho_{\rm MLG}/\rho_{\rm tBLG}$ would be equal to 2. 
The absence of significant Fermi velocity renormalization in tBLG relative to MLG~\cite{supp}, as well as the nearly unchanged electron-TA/LA mode coupling discussed above, support the notion that the two layers are electronically decoupled in large-angle tBLG. This is consistent with the conventional wisdom that momentum mismatch of the rotated Brillouin zones prevents interlayer tunneling~\cite{Polshyn2019}. 
Yet, our computed MLG-to-tBLG resistivity ratio is less than 2 at all temperatures, showing the presence of additional scattering mechanisms that increase the resistivity and \mbox{effectively} couple the two layers. 
\\
\indent
We find that the main mechanism coupling the two layers in large-angle tBLG are $e$-ph interactions due to LB phonons with wave-vectors near K and K' in the tBLG Brillouin zone, which enhance \mbox{intervalley} electron scattering, thereby increasing the resistivity. 
To arrive at this result, we separate the intravalley and intervalley scattering rates in 21.8$^{\circ}$ tBLG by restricting the scattering wave-vector $|\mathbf{q}|$ to appropriate ranges. 
\\
\indent
\mbox{Figure~\ref{fig:trans}(d)-(f)} compare intra- and intervalley carrier scattering at intermediate temperatures between 50$-$225~K. At 225~K, the scattering rate for both intra- and intervalley processes is proportional to the carrier energy relative to the Dirac point, as expected when $e$-ph scattering is controlled by the phase space for energy- and momentum-conserving scattering~\cite{Marco2014}. 
At lower temperatures between $\sim$50$-$200~K, the phonon populations near K and K' decrease, lowering the intervalley scattering rates at the Fermi energy. 
Because the energy of the LB K-valley phonon is approximately \mbox{19 meV} (see Fig.~\ref{fig:gkq}(a)), which corresponds to about 220~K, this mode is thermally excited at such intermediate temperatures and can contribute to the $e$-ph scattering and resistivity down to $\sim$50 K. Thermal excitation of this LB mode, which is clearly absent in MLG, significantly enhances intervalley scattering in large-angle tBLG. 
\\
\indent
Analysis of the $e$-ph interactions in the Wannier basis shows that such intervalley scattering is primarily interlayer (see SM~\cite{supp}), and thus couples the K$_1$ and K$_2$ (and K'$_1$ and K'$_2$) valleys on the two different layers (see Fig.~\ref{fig:tdep}(a)). This mechanism explains the decrease in the MLG-to-tBLG resistivity ratio between 50$-$250 K in Fig.~\ref{fig:trans}(b) and is responsible for the faster-than-linear temperature dependence of the resistivity in large-angle tBLG. 
In addition, due to the slightly lower energy of the K-valley phonon, intervalley scattering in 13.2$^{\circ}$ tBLG becomes important at lower temperatures, and thus the resistivity ratio of 13.2$^{\circ}$ tBLG reaches a minimum at a lower temperature than in 21.8$^{\circ}$ tBLG. 
\\
\indent
In Fig.~\ref{fig:trans}(c), we illustrate the phonon mode with the strongest $e$-ph coupling near the K-point in 21.8$^{\circ}$ tBLG, which gives the strongest contribution to intervalley scattering~\footnote{Of the three degenerate modes at K, the mode shown in Fig. 4(c) is the one with a downward dispersion along the K$-$$\Gamma$ and M$-$K directions. This is the mode with the strongest coupling near K.}. This phonon is a LB mode associated with layer-normal vibrations of the two layers in opposite directions and centered at regions of the moir\'e lattice with local AA stacking. The incipient instability of the AA domains is a unique property of tBLG introduced by the twist-angle degree of freedom, and our calculations reveal its effects on electronic transport. 
\\
\indent
Finally, the intravalley scattering rate in 21.8$^{\circ}$ tBLG is also found to be greater by about 20\% than the intravalley scattering rate in MLG. This difference can be attributed to enhanced intravalley, interlayer scattering mediated by the LB mode at long wavelength (see SM~\cite{supp}). 
Notably, model calculations by Ray et al.~\cite{Ray2016} discussed the effect on transport of LB phonons and their ability to effectively couple the two layers in large-angle tBLG. Their work formulated a model of the low-energy phonons, obtained $e$-ph interactions from tight-binding calculations, and studied interlayer scattering in the same valley, but did not study K-to-K' intervalley scattering. As intervalley processes were neglected, their results are somewhat different from ours and emphasize the role of intravalley scattering. Yet, both studies conclude that low-energy LB phonons effectively couple the two layers in large-angle tBLG. This coupling enhances the resistivity by increasing intra- and (mainly) intervalley scattering. 
\\
\indent 
In summary, we study $e$-ph interactions and electronic transport in large-angle tBLG from first principles, focusing on the role of low-energy phonons. Our approach reveals that acoustic and optical $e$-ph coupling are overall similar in large-angle tBLG and MLG. Yet, $e$-ph interactions due to LB phonons enhance intervalley scattering in tBLG, effectively coupling the two layers and leading to a faster-than-linear temperature dependence of the resistivity. 
Our work sheds light on microscopic interactions in tBLG, paving the way for accurate first-principles studies of electron interactions and transport in tBLG and other systems with twist-angle degree of freedom.\\

\begin{acknowledgments}
This work was primarily supported by the National Science Foundation under Grant No. OAC-2209262. 
This research used resources of the National Energy Research
Scientific Computing Center (NERSC), a DOE Office of Science User Facility
supported by the Office of Science of the U.S. Department of Energy
under Contract No. DE-AC02-05CH11231 using NERSC award
NERSC DDR-ERCAP0026831.
\end{acknowledgments}


\begin{thebibliography}{46}%
\makeatletter
\providecommand \@ifxundefined [1]{%
 \@ifx{#1\undefined}
}%
\providecommand \@ifnum [1]{%
 \ifnum #1\expandafter \@firstoftwo
 \else \expandafter \@secondoftwo
 \fi
}%
\providecommand \@ifx [1]{%
 \ifx #1\expandafter \@firstoftwo
 \else \expandafter \@secondoftwo
 \fi
}%
\providecommand \natexlab [1]{#1}%
\providecommand \enquote  [1]{``#1''}%
\providecommand \bibnamefont  [1]{#1}%
\providecommand \bibfnamefont [1]{#1}%
\providecommand \citenamefont [1]{#1}%
\providecommand \href@noop [0]{\@secondoftwo}%
\providecommand \href [0]{\begingroup \@sanitize@url \@href}%
\providecommand \@href[1]{\@@startlink{#1}\@@href}%
\providecommand \@@href[1]{\endgroup#1\@@endlink}%
\providecommand \@sanitize@url [0]{\catcode `\\12\catcode `\$12\catcode
  `\&12\catcode `\#12\catcode `\^12\catcode `\_12\catcode `\%12\relax}%
\providecommand \@@startlink[1]{}%
\providecommand \@@endlink[0]{}%
\providecommand \url  [0]{\begingroup\@sanitize@url \@url }%
\providecommand \@url [1]{\endgroup\@href {#1}{\urlprefix }}%
\providecommand \urlprefix  [0]{URL }%
\providecommand \Eprint [0]{\href }%
\providecommand \doibase [0]{https://doi.org/}%
\providecommand \selectlanguage [0]{\@gobble}%
\providecommand \bibinfo  [0]{\@secondoftwo}%
\providecommand \bibfield  [0]{\@secondoftwo}%
\providecommand \translation [1]{[#1]}%
\providecommand \BibitemOpen [0]{}%
\providecommand \bibitemStop [0]{}%
\providecommand \bibitemNoStop [0]{.\EOS\space}%
\providecommand \EOS [0]{\spacefactor3000\relax}%
\providecommand \BibitemShut  [1]{\csname bibitem#1\endcsname}%
\let\auto@bib@innerbib\@empty
\bibitem [{\citenamefont {Das~Sarma}\ \emph {et~al.}(2011)\citenamefont
  {Das~Sarma}, \citenamefont {Adam}, \citenamefont {Hwang},\ and\ \citenamefont
  {Rossi}}]{DasSarma2011}%
  \BibitemOpen
  \bibfield  {author} {\bibinfo {author} {\bibfnamefont {S.}~\bibnamefont
  {Das~Sarma}}, \bibinfo {author} {\bibfnamefont {S.}~\bibnamefont {Adam}},
  \bibinfo {author} {\bibfnamefont {E.~H.}\ \bibnamefont {Hwang}},\ and\
  \bibinfo {author} {\bibfnamefont {E.}~\bibnamefont {Rossi}},\ }\bibfield
  {title} {\bibinfo {title} {Electronic transport in two-dimensional
  graphene},\ }\href {https://doi.org/10.1103/RevModPhys.83.407} {\bibfield
  {journal} {\bibinfo  {journal} {Rev. Mod. Phys.}\ }\textbf {\bibinfo {volume}
  {83}},\ \bibinfo {pages} {407} (\bibinfo {year} {2011})}\BibitemShut
  {NoStop}%
\bibitem [{\citenamefont {Hwang}\ and\ \citenamefont
  {Das~Sarma}(2008)}]{Hwang2008}%
  \BibitemOpen
  \bibfield  {author} {\bibinfo {author} {\bibfnamefont {E.~H.}\ \bibnamefont
  {Hwang}}\ and\ \bibinfo {author} {\bibfnamefont {S.}~\bibnamefont
  {Das~Sarma}},\ }\bibfield  {title} {\bibinfo {title} {Acoustic phonon
  scattering limited carrier mobility in two-dimensional extrinsic graphene},\
  }\href {https://doi.org/10.1103/PhysRevB.77.115449} {\bibfield  {journal}
  {\bibinfo  {journal} {Phys. Rev. B}\ }\textbf {\bibinfo {volume} {77}},\
  \bibinfo {pages} {115449} (\bibinfo {year} {2008})}\BibitemShut {NoStop}%
\bibitem [{\citenamefont {Min}\ \emph {et~al.}(2011)\citenamefont {Min},
  \citenamefont {Hwang},\ and\ \citenamefont {Das~Sarma}}]{Min2011}%
  \BibitemOpen
  \bibfield  {author} {\bibinfo {author} {\bibfnamefont {H.}~\bibnamefont
  {Min}}, \bibinfo {author} {\bibfnamefont {E.~H.}\ \bibnamefont {Hwang}},\
  and\ \bibinfo {author} {\bibfnamefont {S.}~\bibnamefont {Das~Sarma}},\
  }\bibfield  {title} {\bibinfo {title} {Chirality-dependent phonon-limited
  resistivity in multiple layers of graphene},\ }\href
  {https://doi.org/10.1103/PhysRevB.83.161404} {\bibfield  {journal} {\bibinfo
  {journal} {Phys. Rev. B}\ }\textbf {\bibinfo {volume} {83}},\ \bibinfo
  {pages} {161404} (\bibinfo {year} {2011})}\BibitemShut {NoStop}%
\bibitem [{\citenamefont {Mariani}\ and\ \citenamefont {von
  Oppen}(2008)}]{Mariani2008}%
  \BibitemOpen
  \bibfield  {author} {\bibinfo {author} {\bibfnamefont {E.}~\bibnamefont
  {Mariani}}\ and\ \bibinfo {author} {\bibfnamefont {F.}~\bibnamefont {von
  Oppen}},\ }\bibfield  {title} {\bibinfo {title} {Flexural phonons in
  free-standing graphene},\ }\href
  {https://doi.org/10.1103/PhysRevLett.100.076801} {\bibfield  {journal}
  {\bibinfo  {journal} {Phys. Rev. Lett.}\ }\textbf {\bibinfo {volume} {100}},\
  \bibinfo {pages} {076801} (\bibinfo {year} {2008})}\BibitemShut {NoStop}%
\bibitem [{\citenamefont {Mariani}\ and\ \citenamefont {von
  Oppen}(2010)}]{Mariani2010}%
  \BibitemOpen
  \bibfield  {author} {\bibinfo {author} {\bibfnamefont {E.}~\bibnamefont
  {Mariani}}\ and\ \bibinfo {author} {\bibfnamefont {F.}~\bibnamefont {von
  Oppen}},\ }\bibfield  {title} {\bibinfo {title} {Temperature-dependent
  resistivity of suspended graphene},\ }\href
  {https://doi.org/10.1103/PhysRevB.82.195403} {\bibfield  {journal} {\bibinfo
  {journal} {Phys. Rev. B}\ }\textbf {\bibinfo {volume} {82}},\ \bibinfo
  {pages} {195403} (\bibinfo {year} {2010})}\BibitemShut {NoStop}%
\bibitem [{\citenamefont {Castro}\ \emph {et~al.}(2010)\citenamefont {Castro},
  \citenamefont {Ochoa}, \citenamefont {Katsnelson}, \citenamefont {Gorbachev},
  \citenamefont {Elias}, \citenamefont {Novoselov}, \citenamefont {Geim},\ and\
  \citenamefont {Guinea}}]{Castro2010}%
  \BibitemOpen
  \bibfield  {author} {\bibinfo {author} {\bibfnamefont {E.~V.}\ \bibnamefont
  {Castro}}, \bibinfo {author} {\bibfnamefont {H.}~\bibnamefont {Ochoa}},
  \bibinfo {author} {\bibfnamefont {M.~I.}\ \bibnamefont {Katsnelson}},
  \bibinfo {author} {\bibfnamefont {R.~V.}\ \bibnamefont {Gorbachev}}, \bibinfo
  {author} {\bibfnamefont {D.~C.}\ \bibnamefont {Elias}}, \bibinfo {author}
  {\bibfnamefont {K.~S.}\ \bibnamefont {Novoselov}}, \bibinfo {author}
  {\bibfnamefont {A.~K.}\ \bibnamefont {Geim}},\ and\ \bibinfo {author}
  {\bibfnamefont {F.}~\bibnamefont {Guinea}},\ }\bibfield  {title} {\bibinfo
  {title} {Limits on charge carrier mobility in suspended graphene due to
  flexural phonons},\ }\href {https://doi.org/10.1103/PhysRevLett.105.266601}
  {\bibfield  {journal} {\bibinfo  {journal} {Phys. Rev. Lett.}\ }\textbf
  {\bibinfo {volume} {105}},\ \bibinfo {pages} {266601} (\bibinfo {year}
  {2010})}\BibitemShut {NoStop}%
\bibitem [{\citenamefont {Borysenko}\ \emph {et~al.}(2010)\citenamefont
  {Borysenko}, \citenamefont {Mullen}, \citenamefont {Barry}, \citenamefont
  {Paul}, \citenamefont {Semenov}, \citenamefont {Zavada}, \citenamefont
  {Nardelli},\ and\ \citenamefont {Kim}}]{Borysenko2010}%
  \BibitemOpen
  \bibfield  {author} {\bibinfo {author} {\bibfnamefont {K.~M.}\ \bibnamefont
  {Borysenko}}, \bibinfo {author} {\bibfnamefont {J.~T.}\ \bibnamefont
  {Mullen}}, \bibinfo {author} {\bibfnamefont {E.~A.}\ \bibnamefont {Barry}},
  \bibinfo {author} {\bibfnamefont {S.}~\bibnamefont {Paul}}, \bibinfo {author}
  {\bibfnamefont {Y.~G.}\ \bibnamefont {Semenov}}, \bibinfo {author}
  {\bibfnamefont {J.~M.}\ \bibnamefont {Zavada}}, \bibinfo {author}
  {\bibfnamefont {M.~B.}\ \bibnamefont {Nardelli}},\ and\ \bibinfo {author}
  {\bibfnamefont {K.~W.}\ \bibnamefont {Kim}},\ }\bibfield  {title} {\bibinfo
  {title} {First-principles analysis of electron-phonon interactions in
  graphene},\ }\href {https://doi.org/10.1103/PhysRevB.81.121412} {\bibfield
  {journal} {\bibinfo  {journal} {Phys. Rev. B}\ }\textbf {\bibinfo {volume}
  {81}},\ \bibinfo {pages} {121412} (\bibinfo {year} {2010})}\BibitemShut
  {NoStop}%
\bibitem [{\citenamefont {Borysenko}\ \emph {et~al.}(2011)\citenamefont
  {Borysenko}, \citenamefont {Mullen}, \citenamefont {Li}, \citenamefont
  {Semenov}, \citenamefont {Zavada}, \citenamefont {Nardelli},\ and\
  \citenamefont {Kim}}]{Borysenko2011}%
  \BibitemOpen
  \bibfield  {author} {\bibinfo {author} {\bibfnamefont {K.~M.}\ \bibnamefont
  {Borysenko}}, \bibinfo {author} {\bibfnamefont {J.~T.}\ \bibnamefont
  {Mullen}}, \bibinfo {author} {\bibfnamefont {X.}~\bibnamefont {Li}}, \bibinfo
  {author} {\bibfnamefont {Y.~G.}\ \bibnamefont {Semenov}}, \bibinfo {author}
  {\bibfnamefont {J.~M.}\ \bibnamefont {Zavada}}, \bibinfo {author}
  {\bibfnamefont {M.~B.}\ \bibnamefont {Nardelli}},\ and\ \bibinfo {author}
  {\bibfnamefont {K.~W.}\ \bibnamefont {Kim}},\ }\bibfield  {title} {\bibinfo
  {title} {Electron-phonon interactions in bilayer graphene},\ }\href
  {https://doi.org/10.1103/PhysRevB.83.161402} {\bibfield  {journal} {\bibinfo
  {journal} {Phys. Rev. B}\ }\textbf {\bibinfo {volume} {83}},\ \bibinfo
  {pages} {161402} (\bibinfo {year} {2011})}\BibitemShut {NoStop}%
\bibitem [{\citenamefont {Li}\ \emph {et~al.}(2011)\citenamefont {Li},
  \citenamefont {Borysenko}, \citenamefont {Nardelli},\ and\ \citenamefont
  {Kim}}]{Li2011}%
  \BibitemOpen
  \bibfield  {author} {\bibinfo {author} {\bibfnamefont {X.}~\bibnamefont
  {Li}}, \bibinfo {author} {\bibfnamefont {K.~M.}\ \bibnamefont {Borysenko}},
  \bibinfo {author} {\bibfnamefont {M.~B.}\ \bibnamefont {Nardelli}},\ and\
  \bibinfo {author} {\bibfnamefont {K.~W.}\ \bibnamefont {Kim}},\ }\bibfield
  {title} {\bibinfo {title} {Electron transport properties of bilayer
  graphene},\ }\href {https://doi.org/10.1103/PhysRevB.84.195453} {\bibfield
  {journal} {\bibinfo  {journal} {Phys. Rev. B}\ }\textbf {\bibinfo {volume}
  {84}},\ \bibinfo {pages} {195453} (\bibinfo {year} {2011})}\BibitemShut
  {NoStop}%
\bibitem [{\citenamefont {Kaasbjerg}\ \emph {et~al.}(2012)\citenamefont
  {Kaasbjerg}, \citenamefont {Thygesen},\ and\ \citenamefont
  {Jacobsen}}]{Kaasbjerg2012}%
  \BibitemOpen
  \bibfield  {author} {\bibinfo {author} {\bibfnamefont {K.}~\bibnamefont
  {Kaasbjerg}}, \bibinfo {author} {\bibfnamefont {K.~S.}\ \bibnamefont
  {Thygesen}},\ and\ \bibinfo {author} {\bibfnamefont {K.~W.}\ \bibnamefont
  {Jacobsen}},\ }\bibfield  {title} {\bibinfo {title} {Unraveling the acoustic
  electron-phonon interaction in graphene},\ }\href
  {https://doi.org/10.1103/PhysRevB.85.165440} {\bibfield  {journal} {\bibinfo
  {journal} {Phys. Rev. B}\ }\textbf {\bibinfo {volume} {85}},\ \bibinfo
  {pages} {165440} (\bibinfo {year} {2012})}\BibitemShut {NoStop}%
\bibitem [{\citenamefont {Park}\ \emph {et~al.}(2014)\citenamefont {Park},
  \citenamefont {Bonini}, \citenamefont {Sohier}, \citenamefont {Samsonidze},
  \citenamefont {Kozinsky}, \citenamefont {Calandra}, \citenamefont {Mauri},\
  and\ \citenamefont {Marzari}}]{Park2014}%
  \BibitemOpen
  \bibfield  {author} {\bibinfo {author} {\bibfnamefont {C.~H.}\ \bibnamefont
  {Park}}, \bibinfo {author} {\bibfnamefont {N.}~\bibnamefont {Bonini}},
  \bibinfo {author} {\bibfnamefont {T.}~\bibnamefont {Sohier}}, \bibinfo
  {author} {\bibfnamefont {G.}~\bibnamefont {Samsonidze}}, \bibinfo {author}
  {\bibfnamefont {B.}~\bibnamefont {Kozinsky}}, \bibinfo {author}
  {\bibfnamefont {M.}~\bibnamefont {Calandra}}, \bibinfo {author}
  {\bibfnamefont {F.}~\bibnamefont {Mauri}},\ and\ \bibinfo {author}
  {\bibfnamefont {N.}~\bibnamefont {Marzari}},\ }\bibfield  {title} {\bibinfo
  {title} {{Electron-phonon interactions and the intrinsic electrical
  resistivity of graphene}},\ }\href {https://doi.org/10.1021/nl402696q}
  {\bibfield  {journal} {\bibinfo  {journal} {Nano Lett.}\ }\textbf {\bibinfo
  {volume} {14}},\ \bibinfo {pages} {1113} (\bibinfo {year}
  {2014})}\BibitemShut {NoStop}%
\bibitem [{\citenamefont {Sohier}\ \emph {et~al.}(2014)\citenamefont {Sohier},
  \citenamefont {Calandra}, \citenamefont {Park}, \citenamefont {Bonini},
  \citenamefont {Marzari},\ and\ \citenamefont {Mauri}}]{Sohier2014}%
  \BibitemOpen
  \bibfield  {author} {\bibinfo {author} {\bibfnamefont {T.}~\bibnamefont
  {Sohier}}, \bibinfo {author} {\bibfnamefont {M.}~\bibnamefont {Calandra}},
  \bibinfo {author} {\bibfnamefont {C.-H.}\ \bibnamefont {Park}}, \bibinfo
  {author} {\bibfnamefont {N.}~\bibnamefont {Bonini}}, \bibinfo {author}
  {\bibfnamefont {N.}~\bibnamefont {Marzari}},\ and\ \bibinfo {author}
  {\bibfnamefont {F.}~\bibnamefont {Mauri}},\ }\bibfield  {title} {\bibinfo
  {title} {Phonon-limited resistivity of graphene by first-principles
  calculations: Electron-phonon interactions, strain-induced gauge field, and
  boltzmann equation},\ }\href {https://doi.org/10.1103/PhysRevB.90.125414}
  {\bibfield  {journal} {\bibinfo  {journal} {Phys. Rev. B}\ }\textbf {\bibinfo
  {volume} {90}},\ \bibinfo {pages} {125414} (\bibinfo {year}
  {2014})}\BibitemShut {NoStop}%
\bibitem [{\citenamefont {Cao}\ \emph {et~al.}(2018{\natexlab{a}})\citenamefont
  {Cao}, \citenamefont {Fatemi}, \citenamefont {Demir}, \citenamefont {Fang},
  \citenamefont {Tomarken}, \citenamefont {Luo}, \citenamefont
  {Sanchez-Yamagishi}, \citenamefont {Watanabe}, \citenamefont {Taniguchi},
  \citenamefont {Kaxiras}, \citenamefont {Ashoori},\ and\ \citenamefont
  {Jarillo-Herrero}}]{Cao2018}%
  \BibitemOpen
  \bibfield  {author} {\bibinfo {author} {\bibfnamefont {Y.}~\bibnamefont
  {Cao}}, \bibinfo {author} {\bibfnamefont {V.}~\bibnamefont {Fatemi}},
  \bibinfo {author} {\bibfnamefont {A.}~\bibnamefont {Demir}}, \bibinfo
  {author} {\bibfnamefont {S.}~\bibnamefont {Fang}}, \bibinfo {author}
  {\bibfnamefont {S.~L.}\ \bibnamefont {Tomarken}}, \bibinfo {author}
  {\bibfnamefont {J.~Y.}\ \bibnamefont {Luo}}, \bibinfo {author} {\bibfnamefont
  {J.~D.}\ \bibnamefont {Sanchez-Yamagishi}}, \bibinfo {author} {\bibfnamefont
  {K.}~\bibnamefont {Watanabe}}, \bibinfo {author} {\bibfnamefont
  {T.}~\bibnamefont {Taniguchi}}, \bibinfo {author} {\bibfnamefont
  {E.}~\bibnamefont {Kaxiras}}, \bibinfo {author} {\bibfnamefont {R.~C.}\
  \bibnamefont {Ashoori}},\ and\ \bibinfo {author} {\bibfnamefont
  {P.}~\bibnamefont {Jarillo-Herrero}},\ }\bibfield  {title} {\bibinfo {title}
  {{Correlated insulator behaviour at half-filling in magic-angle graphene
  superlattices}},\ }\href {https://doi.org/10.1038/nature26154} {\bibfield
  {journal} {\bibinfo  {journal} {Nature}\ }\textbf {\bibinfo {volume} {556}},\
  \bibinfo {pages} {80} (\bibinfo {year} {2018}{\natexlab{a}})}\BibitemShut
  {NoStop}%
\bibitem [{\citenamefont {Cao}\ \emph {et~al.}(2018{\natexlab{b}})\citenamefont
  {Cao}, \citenamefont {Fatemi}, \citenamefont {Fang}, \citenamefont
  {Watanabe}, \citenamefont {Taniguchi}, \citenamefont {Kaxiras},\ and\
  \citenamefont {Jarillo-Herrero}}]{Cao2018a}%
  \BibitemOpen
  \bibfield  {author} {\bibinfo {author} {\bibfnamefont {Y.}~\bibnamefont
  {Cao}}, \bibinfo {author} {\bibfnamefont {V.}~\bibnamefont {Fatemi}},
  \bibinfo {author} {\bibfnamefont {S.}~\bibnamefont {Fang}}, \bibinfo {author}
  {\bibfnamefont {K.}~\bibnamefont {Watanabe}}, \bibinfo {author}
  {\bibfnamefont {T.}~\bibnamefont {Taniguchi}}, \bibinfo {author}
  {\bibfnamefont {E.}~\bibnamefont {Kaxiras}},\ and\ \bibinfo {author}
  {\bibfnamefont {P.}~\bibnamefont {Jarillo-Herrero}},\ }\bibfield  {title}
  {\bibinfo {title} {{Unconventional superconductivity in magic-angle graphene
  superlattices}},\ }\href {https://doi.org/10.1038/nature26160} {\bibfield
  {journal} {\bibinfo  {journal} {Nature}\ }\textbf {\bibinfo {volume} {556}},\
  \bibinfo {pages} {43} (\bibinfo {year} {2018}{\natexlab{b}})}\BibitemShut
  {NoStop}%
\bibitem [{\citenamefont {Choi}\ and\ \citenamefont {Choi}(2018)}]{Choi2018}%
  \BibitemOpen
  \bibfield  {author} {\bibinfo {author} {\bibfnamefont {Y.~W.}\ \bibnamefont
  {Choi}}\ and\ \bibinfo {author} {\bibfnamefont {H.~J.}\ \bibnamefont
  {Choi}},\ }\bibfield  {title} {\bibinfo {title} {Strong electron-phonon
  coupling, electron-hole asymmetry, and nonadiabaticity in magic-angle twisted
  bilayer graphene},\ }\href {https://doi.org/10.1103/PhysRevB.98.241412}
  {\bibfield  {journal} {\bibinfo  {journal} {Phys. Rev. B}\ }\textbf {\bibinfo
  {volume} {98}},\ \bibinfo {pages} {241412} (\bibinfo {year}
  {2018})}\BibitemShut {NoStop}%
\bibitem [{\citenamefont {Peltonen}\ \emph {et~al.}(2018)\citenamefont
  {Peltonen}, \citenamefont {Ojaj\"arvi},\ and\ \citenamefont
  {Heikkil\"a}}]{Peltonen2018}%
  \BibitemOpen
  \bibfield  {author} {\bibinfo {author} {\bibfnamefont {T.~J.}\ \bibnamefont
  {Peltonen}}, \bibinfo {author} {\bibfnamefont {R.}~\bibnamefont
  {Ojaj\"arvi}},\ and\ \bibinfo {author} {\bibfnamefont {T.~T.}\ \bibnamefont
  {Heikkil\"a}},\ }\bibfield  {title} {\bibinfo {title} {Mean-field theory for
  superconductivity in twisted bilayer graphene},\ }\href
  {https://doi.org/10.1103/PhysRevB.98.220504} {\bibfield  {journal} {\bibinfo
  {journal} {Phys. Rev. B}\ }\textbf {\bibinfo {volume} {98}},\ \bibinfo
  {pages} {220504} (\bibinfo {year} {2018})}\BibitemShut {NoStop}%
\bibitem [{\citenamefont {Wu}\ \emph {et~al.}(2018)\citenamefont {Wu},
  \citenamefont {MacDonald},\ and\ \citenamefont {Martin}}]{Wu2018}%
  \BibitemOpen
  \bibfield  {author} {\bibinfo {author} {\bibfnamefont {F.}~\bibnamefont
  {Wu}}, \bibinfo {author} {\bibfnamefont {A.~H.}\ \bibnamefont {MacDonald}},\
  and\ \bibinfo {author} {\bibfnamefont {I.}~\bibnamefont {Martin}},\
  }\bibfield  {title} {\bibinfo {title} {Theory of phonon-mediated
  superconductivity in twisted bilayer graphene},\ }\href
  {https://doi.org/10.1103/PhysRevLett.121.257001} {\bibfield  {journal}
  {\bibinfo  {journal} {Phys. Rev. Lett.}\ }\textbf {\bibinfo {volume} {121}},\
  \bibinfo {pages} {257001} (\bibinfo {year} {2018})}\BibitemShut {NoStop}%
\bibitem [{\citenamefont {Lian}\ \emph {et~al.}(2019)\citenamefont {Lian},
  \citenamefont {Wang},\ and\ \citenamefont {Bernevig}}]{Lian2019}%
  \BibitemOpen
  \bibfield  {author} {\bibinfo {author} {\bibfnamefont {B.}~\bibnamefont
  {Lian}}, \bibinfo {author} {\bibfnamefont {Z.}~\bibnamefont {Wang}},\ and\
  \bibinfo {author} {\bibfnamefont {B.~A.}\ \bibnamefont {Bernevig}},\
  }\bibfield  {title} {\bibinfo {title} {Twisted bilayer graphene: A
  phonon-driven superconductor},\ }\href
  {https://doi.org/10.1103/PhysRevLett.122.257002} {\bibfield  {journal}
  {\bibinfo  {journal} {Phys. Rev. Lett.}\ }\textbf {\bibinfo {volume} {122}},\
  \bibinfo {pages} {257002} (\bibinfo {year} {2019})}\BibitemShut {NoStop}%
\bibitem [{\citenamefont {Andrei}\ and\ \citenamefont
  {MacDonald}(2020)}]{Andrei2020}%
  \BibitemOpen
  \bibfield  {author} {\bibinfo {author} {\bibfnamefont {E.~Y.}\ \bibnamefont
  {Andrei}}\ and\ \bibinfo {author} {\bibfnamefont {A.~H.}\ \bibnamefont
  {MacDonald}},\ }\bibfield  {title} {\bibinfo {title} {{Graphene bilayers with
  a twist}},\ }\href {https://doi.org/10.1038/s41563-020-00840-0} {\bibfield
  {journal} {\bibinfo  {journal} {Nat. Mater.}\ }\textbf {\bibinfo {volume}
  {19}},\ \bibinfo {pages} {1265} (\bibinfo {year} {2020})}\BibitemShut
  {NoStop}%
\bibitem [{\citenamefont {Polshyn}\ \emph {et~al.}(2019)\citenamefont
  {Polshyn}, \citenamefont {Yankowitz}, \citenamefont {Chen}, \citenamefont
  {Zhang}, \citenamefont {Watanabe}, \citenamefont {Taniguchi}, \citenamefont
  {Dean},\ and\ \citenamefont {Young}}]{Polshyn2019}%
  \BibitemOpen
  \bibfield  {author} {\bibinfo {author} {\bibfnamefont {H.}~\bibnamefont
  {Polshyn}}, \bibinfo {author} {\bibfnamefont {M.}~\bibnamefont {Yankowitz}},
  \bibinfo {author} {\bibfnamefont {S.}~\bibnamefont {Chen}}, \bibinfo {author}
  {\bibfnamefont {Y.}~\bibnamefont {Zhang}}, \bibinfo {author} {\bibfnamefont
  {K.}~\bibnamefont {Watanabe}}, \bibinfo {author} {\bibfnamefont
  {T.}~\bibnamefont {Taniguchi}}, \bibinfo {author} {\bibfnamefont {C.~R.}\
  \bibnamefont {Dean}},\ and\ \bibinfo {author} {\bibfnamefont {A.~F.}\
  \bibnamefont {Young}},\ }\bibfield  {title} {\bibinfo {title} {{Large
  linear-in-temperature resistivity in twisted bilayer graphene}},\ }\href
  {https://doi.org/10.1038/s41567-019-0596-3} {\bibfield  {journal} {\bibinfo
  {journal} {Nat. Phys.}\ }\textbf {\bibinfo {volume} {15}},\ \bibinfo {pages}
  {1011} (\bibinfo {year} {2019})}\BibitemShut {NoStop}%
\bibitem [{\citenamefont {Ray}\ \emph {et~al.}(2016)\citenamefont {Ray},
  \citenamefont {Fleischmann}, \citenamefont {Weckbecker}, \citenamefont
  {Sharma}, \citenamefont {Pankratov},\ and\ \citenamefont
  {Shallcross}}]{Ray2016}%
  \BibitemOpen
  \bibfield  {author} {\bibinfo {author} {\bibfnamefont {N.}~\bibnamefont
  {Ray}}, \bibinfo {author} {\bibfnamefont {M.}~\bibnamefont {Fleischmann}},
  \bibinfo {author} {\bibfnamefont {D.}~\bibnamefont {Weckbecker}}, \bibinfo
  {author} {\bibfnamefont {S.}~\bibnamefont {Sharma}}, \bibinfo {author}
  {\bibfnamefont {O.}~\bibnamefont {Pankratov}},\ and\ \bibinfo {author}
  {\bibfnamefont {S.}~\bibnamefont {Shallcross}},\ }\bibfield  {title}
  {\bibinfo {title} {Electron-phonon scattering and in-plane electric
  conductivity in twisted bilayer graphene},\ }\href
  {https://doi.org/10.1103/PhysRevB.94.245403} {\bibfield  {journal} {\bibinfo
  {journal} {Phys. Rev. B}\ }\textbf {\bibinfo {volume} {94}},\ \bibinfo
  {pages} {245403} (\bibinfo {year} {2016})}\BibitemShut {NoStop}%
\bibitem [{\citenamefont {Ochoa}(2019)}]{Ochoa2019}%
  \BibitemOpen
  \bibfield  {author} {\bibinfo {author} {\bibfnamefont {H.}~\bibnamefont
  {Ochoa}},\ }\bibfield  {title} {\bibinfo {title} {Moir\'e-pattern
  fluctuations and electron-phason coupling in twisted bilayer graphene},\
  }\href {https://doi.org/10.1103/PhysRevB.100.155426} {\bibfield  {journal}
  {\bibinfo  {journal} {Phys. Rev. B}\ }\textbf {\bibinfo {volume} {100}},\
  \bibinfo {pages} {155426} (\bibinfo {year} {2019})}\BibitemShut {NoStop}%
\bibitem [{\citenamefont {Wu}\ \emph {et~al.}(2019)\citenamefont {Wu},
  \citenamefont {Hwang},\ and\ \citenamefont {Das~Sarma}}]{Wu2019}%
  \BibitemOpen
  \bibfield  {author} {\bibinfo {author} {\bibfnamefont {F.}~\bibnamefont
  {Wu}}, \bibinfo {author} {\bibfnamefont {E.}~\bibnamefont {Hwang}},\ and\
  \bibinfo {author} {\bibfnamefont {S.}~\bibnamefont {Das~Sarma}},\ }\bibfield
  {title} {\bibinfo {title} {Phonon-induced giant linear-in-{$T$} resistivity
  in magic angle twisted bilayer graphene: Ordinary strangeness and exotic
  superconductivity},\ }\href {https://doi.org/10.1103/PhysRevB.99.165112}
  {\bibfield  {journal} {\bibinfo  {journal} {Phys. Rev. B}\ }\textbf {\bibinfo
  {volume} {99}},\ \bibinfo {pages} {165112} (\bibinfo {year}
  {2019})}\BibitemShut {NoStop}%
\bibitem [{\citenamefont {Li}\ \emph {et~al.}(2020)\citenamefont {Li},
  \citenamefont {Wu},\ and\ \citenamefont {Das~Sarma}}]{Li2020}%
  \BibitemOpen
  \bibfield  {author} {\bibinfo {author} {\bibfnamefont {X.}~\bibnamefont
  {Li}}, \bibinfo {author} {\bibfnamefont {F.}~\bibnamefont {Wu}},\ and\
  \bibinfo {author} {\bibfnamefont {S.}~\bibnamefont {Das~Sarma}},\ }\bibfield
  {title} {\bibinfo {title} {Phonon scattering induced carrier resistivity in
  twisted double-bilayer graphene},\ }\href
  {https://doi.org/10.1103/PhysRevB.101.245436} {\bibfield  {journal} {\bibinfo
   {journal} {Phys. Rev. B}\ }\textbf {\bibinfo {volume} {101}},\ \bibinfo
  {pages} {245436} (\bibinfo {year} {2020})}\BibitemShut {NoStop}%
\bibitem [{\citenamefont {Nam}\ and\ \citenamefont
  {Koshino}(2017)}]{Koshino2017}%
  \BibitemOpen
  \bibfield  {author} {\bibinfo {author} {\bibfnamefont {N.~N.~T.}\
  \bibnamefont {Nam}}\ and\ \bibinfo {author} {\bibfnamefont {M.}~\bibnamefont
  {Koshino}},\ }\bibfield  {title} {\bibinfo {title} {Lattice relaxation and
  energy band modulation in twisted bilayer graphene},\ }\href
  {https://link.aps.org/doi/10.1103/PhysRevB.96.075311} {\bibfield  {journal}
  {\bibinfo  {journal} {Phys. Rev. B}\ }\textbf {\bibinfo {volume} {96}},\
  \bibinfo {pages} {075311} (\bibinfo {year} {2017})}\BibitemShut {NoStop}%
\bibitem [{\citenamefont {Carr}\ \emph {et~al.}(2019)\citenamefont {Carr},
  \citenamefont {Fang}, \citenamefont {Zhu},\ and\ \citenamefont
  {Kaxiras}}]{Kaxiras2019}%
  \BibitemOpen
  \bibfield  {author} {\bibinfo {author} {\bibfnamefont {S.}~\bibnamefont
  {Carr}}, \bibinfo {author} {\bibfnamefont {S.}~\bibnamefont {Fang}}, \bibinfo
  {author} {\bibfnamefont {Z.}~\bibnamefont {Zhu}},\ and\ \bibinfo {author}
  {\bibfnamefont {E.}~\bibnamefont {Kaxiras}},\ }\bibfield  {title} {\bibinfo
  {title} {Exact continuum model for low-energy electronic states of twisted
  bilayer graphene},\ }\href
  {https://link.aps.org/doi/10.1103/PhysRevResearch.1.013001} {\bibfield
  {journal} {\bibinfo  {journal} {Phys. Rev. Res.}\ }\textbf {\bibinfo {volume}
  {1}},\ \bibinfo {pages} {013001} (\bibinfo {year} {2019})}\BibitemShut
  {NoStop}%
\bibitem [{\citenamefont {Pathak}\ \emph {et~al.}(2022)\citenamefont {Pathak},
  \citenamefont {Rakib}, \citenamefont {Hou}, \citenamefont {Nevidomskyy},
  \citenamefont {Ertekin}, \citenamefont {Johnson},\ and\ \citenamefont
  {Wagner}}]{Wagner2022}%
  \BibitemOpen
  \bibfield  {author} {\bibinfo {author} {\bibfnamefont {S.}~\bibnamefont
  {Pathak}}, \bibinfo {author} {\bibfnamefont {T.}~\bibnamefont {Rakib}},
  \bibinfo {author} {\bibfnamefont {R.}~\bibnamefont {Hou}}, \bibinfo {author}
  {\bibfnamefont {A.}~\bibnamefont {Nevidomskyy}}, \bibinfo {author}
  {\bibfnamefont {E.}~\bibnamefont {Ertekin}}, \bibinfo {author} {\bibfnamefont
  {H.~T.}\ \bibnamefont {Johnson}},\ and\ \bibinfo {author} {\bibfnamefont
  {L.~K.}\ \bibnamefont {Wagner}},\ }\bibfield  {title} {\bibinfo {title}
  {Accurate tight-binding model for twisted bilayer graphene describes
  topological flat bands without geometric relaxation},\ }\href
  {https://link.aps.org/doi/10.1103/PhysRevB.105.115141} {\bibfield  {journal}
  {\bibinfo  {journal} {Phys. Rev. B}\ }\textbf {\bibinfo {volume} {105}},\
  \bibinfo {pages} {115141} (\bibinfo {year} {2022})}\BibitemShut {NoStop}%
\bibitem [{\citenamefont {Mele}(2010)}]{Mele2010}%
  \BibitemOpen
  \bibfield  {author} {\bibinfo {author} {\bibfnamefont {E.~J.}\ \bibnamefont
  {Mele}},\ }\bibfield  {title} {\bibinfo {title} {Commensuration and
  interlayer coherence in twisted bilayer graphene},\ }\href
  {https://doi.org/10.1103/PhysRevB.81.161405} {\bibfield  {journal} {\bibinfo
  {journal} {Phys. Rev. B}\ }\textbf {\bibinfo {volume} {81}},\ \bibinfo
  {pages} {161405} (\bibinfo {year} {2010})}\BibitemShut {NoStop}%
\bibitem [{\citenamefont {Moon}\ and\ \citenamefont
  {Koshino}(2013)}]{Moon2013}%
  \BibitemOpen
  \bibfield  {author} {\bibinfo {author} {\bibfnamefont {P.}~\bibnamefont
  {Moon}}\ and\ \bibinfo {author} {\bibfnamefont {M.}~\bibnamefont {Koshino}},\
  }\bibfield  {title} {\bibinfo {title} {Optical absorption in twisted bilayer
  graphene},\ }\href {https://doi.org/10.1103/PhysRevB.87.205404} {\bibfield
  {journal} {\bibinfo  {journal} {Phys. Rev. B}\ }\textbf {\bibinfo {volume}
  {87}},\ \bibinfo {pages} {205404} (\bibinfo {year} {2013})}\BibitemShut
  {NoStop}%
\bibitem [{\citenamefont {Giannozzi}\ \emph {et~al.}(2009)\citenamefont
  {Giannozzi}, \citenamefont {Baroni}, \citenamefont {Bonini}, \citenamefont
  {Calandra}, \citenamefont {Car}, \citenamefont {Cavazzoni}, \citenamefont
  {Ceresoli}, \citenamefont {Chiarotti}, \citenamefont {Cococcioni},
  \citenamefont {Dabo}, \citenamefont {Dal~Corso}, \citenamefont
  {de~Gironcoli}, \citenamefont {Fabris}, \citenamefont {Fratesi},
  \citenamefont {Gebauer}, \citenamefont {Gerstmann}, \citenamefont
  {Gougoussis}, \citenamefont {Kokalj}, \citenamefont {Lazzeri}, \citenamefont
  {Martin-Samos}, \citenamefont {Marzari}, \citenamefont {Mauri}, \citenamefont
  {Mazzarello}, \citenamefont {Paolini}, \citenamefont {Pasquarello},
  \citenamefont {Paulatto}, \citenamefont {Sbraccia}, \citenamefont {Scandolo},
  \citenamefont {Sclauzero}, \citenamefont {Seitsonen}, \citenamefont
  {Smogunov}, \citenamefont {Umari},\ and\ \citenamefont
  {Wentzcovitch}}]{giannozzi_quantum_2009}%
  \BibitemOpen
  \bibfield  {author} {\bibinfo {author} {\bibfnamefont {P.}~\bibnamefont
  {Giannozzi}}, \bibinfo {author} {\bibfnamefont {S.}~\bibnamefont {Baroni}},
  \bibinfo {author} {\bibfnamefont {N.}~\bibnamefont {Bonini}}, \bibinfo
  {author} {\bibfnamefont {M.}~\bibnamefont {Calandra}}, \bibinfo {author}
  {\bibfnamefont {R.}~\bibnamefont {Car}}, \bibinfo {author} {\bibfnamefont
  {C.}~\bibnamefont {Cavazzoni}}, \bibinfo {author} {\bibfnamefont
  {D.}~\bibnamefont {Ceresoli}}, \bibinfo {author} {\bibfnamefont {G.~L.}\
  \bibnamefont {Chiarotti}}, \bibinfo {author} {\bibfnamefont {M.}~\bibnamefont
  {Cococcioni}}, \bibinfo {author} {\bibfnamefont {I.}~\bibnamefont {Dabo}},
  \bibinfo {author} {\bibfnamefont {A.}~\bibnamefont {Dal~Corso}}, \bibinfo
  {author} {\bibfnamefont {S.}~\bibnamefont {de~Gironcoli}}, \bibinfo {author}
  {\bibfnamefont {S.}~\bibnamefont {Fabris}}, \bibinfo {author} {\bibfnamefont
  {G.}~\bibnamefont {Fratesi}}, \bibinfo {author} {\bibfnamefont
  {R.}~\bibnamefont {Gebauer}}, \bibinfo {author} {\bibfnamefont
  {U.}~\bibnamefont {Gerstmann}}, \bibinfo {author} {\bibfnamefont
  {C.}~\bibnamefont {Gougoussis}}, \bibinfo {author} {\bibfnamefont
  {A.}~\bibnamefont {Kokalj}}, \bibinfo {author} {\bibfnamefont
  {M.}~\bibnamefont {Lazzeri}}, \bibinfo {author} {\bibfnamefont
  {L.}~\bibnamefont {Martin-Samos}}, \bibinfo {author} {\bibfnamefont
  {N.}~\bibnamefont {Marzari}}, \bibinfo {author} {\bibfnamefont
  {F.}~\bibnamefont {Mauri}}, \bibinfo {author} {\bibfnamefont
  {R.}~\bibnamefont {Mazzarello}}, \bibinfo {author} {\bibfnamefont
  {S.}~\bibnamefont {Paolini}}, \bibinfo {author} {\bibfnamefont
  {A.}~\bibnamefont {Pasquarello}}, \bibinfo {author} {\bibfnamefont
  {L.}~\bibnamefont {Paulatto}}, \bibinfo {author} {\bibfnamefont
  {C.}~\bibnamefont {Sbraccia}}, \bibinfo {author} {\bibfnamefont
  {S.}~\bibnamefont {Scandolo}}, \bibinfo {author} {\bibfnamefont
  {G.}~\bibnamefont {Sclauzero}}, \bibinfo {author} {\bibfnamefont {A.~P.}\
  \bibnamefont {Seitsonen}}, \bibinfo {author} {\bibfnamefont {A.}~\bibnamefont
  {Smogunov}}, \bibinfo {author} {\bibfnamefont {P.}~\bibnamefont {Umari}},\
  and\ \bibinfo {author} {\bibfnamefont {R.~M.}\ \bibnamefont {Wentzcovitch}},\
  }\bibfield  {title} {\bibinfo {title} {{QUANTUM} {ESPRESSO}: a modular and
  open-source software project for quantum simulations of materials},\ }\href
  {https://doi.org/10.1088/0953-8984/21/39/395502} {\bibfield  {journal}
  {\bibinfo  {journal} {J. Phys. Condens. Matter}\ }\textbf {\bibinfo {volume}
  {21}},\ \bibinfo {pages} {395502} (\bibinfo {year} {2009})}\BibitemShut
  {NoStop}%
\bibitem [{\citenamefont {Giannozzi}\ \emph {et~al.}(2017)\citenamefont
  {Giannozzi}, \citenamefont {Andreussi}, \citenamefont {Brumme}, \citenamefont
  {Bunau}, \citenamefont {Buongiorno~Nardelli}, \citenamefont {Calandra},
  \citenamefont {Car}, \citenamefont {Cavazzoni}, \citenamefont {Ceresoli},
  \citenamefont {Cococcioni}, \citenamefont {Colonna}, \citenamefont
  {Carnimeo}, \citenamefont {Dal~Corso}, \citenamefont {de~Gironcoli},
  \citenamefont {Delugas}, \citenamefont {DiStasio}, \citenamefont {Ferretti},
  \citenamefont {Floris}, \citenamefont {Fratesi}, \citenamefont {Fugallo},
  \citenamefont {Gebauer}, \citenamefont {Gerstmann}, \citenamefont {Giustino},
  \citenamefont {Gorni}, \citenamefont {Jia}, \citenamefont {Kawamura},
  \citenamefont {Ko}, \citenamefont {Kokalj}, \citenamefont {Küçükbenli},
  \citenamefont {Lazzeri}, \citenamefont {Marsili}, \citenamefont {Marzari},
  \citenamefont {Mauri}, \citenamefont {Nguyen}, \citenamefont {Nguyen},
  \citenamefont {Otero-de-la Roza}, \citenamefont {Paulatto}, \citenamefont
  {Poncé}, \citenamefont {Rocca}, \citenamefont {Sabatini}, \citenamefont
  {Santra}, \citenamefont {Schlipf}, \citenamefont {Seitsonen}, \citenamefont
  {Smogunov}, \citenamefont {Timrov}, \citenamefont {Thonhauser}, \citenamefont
  {Umari}, \citenamefont {Vast}, \citenamefont {Wu},\ and\ \citenamefont
  {Baroni}}]{giannozzi_advanced_2017}%
  \BibitemOpen
  \bibfield  {author} {\bibinfo {author} {\bibfnamefont {P.}~\bibnamefont
  {Giannozzi}}, \bibinfo {author} {\bibfnamefont {O.}~\bibnamefont
  {Andreussi}}, \bibinfo {author} {\bibfnamefont {T.}~\bibnamefont {Brumme}},
  \bibinfo {author} {\bibfnamefont {O.}~\bibnamefont {Bunau}}, \bibinfo
  {author} {\bibfnamefont {M.}~\bibnamefont {Buongiorno~Nardelli}}, \bibinfo
  {author} {\bibfnamefont {M.}~\bibnamefont {Calandra}}, \bibinfo {author}
  {\bibfnamefont {R.}~\bibnamefont {Car}}, \bibinfo {author} {\bibfnamefont
  {C.}~\bibnamefont {Cavazzoni}}, \bibinfo {author} {\bibfnamefont
  {D.}~\bibnamefont {Ceresoli}}, \bibinfo {author} {\bibfnamefont
  {M.}~\bibnamefont {Cococcioni}}, \bibinfo {author} {\bibfnamefont
  {N.}~\bibnamefont {Colonna}}, \bibinfo {author} {\bibfnamefont
  {I.}~\bibnamefont {Carnimeo}}, \bibinfo {author} {\bibfnamefont
  {A.}~\bibnamefont {Dal~Corso}}, \bibinfo {author} {\bibfnamefont
  {S.}~\bibnamefont {de~Gironcoli}}, \bibinfo {author} {\bibfnamefont
  {P.}~\bibnamefont {Delugas}}, \bibinfo {author} {\bibfnamefont {R.~A.}\
  \bibnamefont {DiStasio}}, \bibinfo {author} {\bibfnamefont {A.}~\bibnamefont
  {Ferretti}}, \bibinfo {author} {\bibfnamefont {A.}~\bibnamefont {Floris}},
  \bibinfo {author} {\bibfnamefont {G.}~\bibnamefont {Fratesi}}, \bibinfo
  {author} {\bibfnamefont {G.}~\bibnamefont {Fugallo}}, \bibinfo {author}
  {\bibfnamefont {R.}~\bibnamefont {Gebauer}}, \bibinfo {author} {\bibfnamefont
  {U.}~\bibnamefont {Gerstmann}}, \bibinfo {author} {\bibfnamefont
  {F.}~\bibnamefont {Giustino}}, \bibinfo {author} {\bibfnamefont
  {T.}~\bibnamefont {Gorni}}, \bibinfo {author} {\bibfnamefont
  {J.}~\bibnamefont {Jia}}, \bibinfo {author} {\bibfnamefont {M.}~\bibnamefont
  {Kawamura}}, \bibinfo {author} {\bibfnamefont {H.-Y.}\ \bibnamefont {Ko}},
  \bibinfo {author} {\bibfnamefont {A.}~\bibnamefont {Kokalj}}, \bibinfo
  {author} {\bibfnamefont {E.}~\bibnamefont {Küçükbenli}}, \bibinfo {author}
  {\bibfnamefont {M.}~\bibnamefont {Lazzeri}}, \bibinfo {author} {\bibfnamefont
  {M.}~\bibnamefont {Marsili}}, \bibinfo {author} {\bibfnamefont
  {N.}~\bibnamefont {Marzari}}, \bibinfo {author} {\bibfnamefont
  {F.}~\bibnamefont {Mauri}}, \bibinfo {author} {\bibfnamefont {N.~L.}\
  \bibnamefont {Nguyen}}, \bibinfo {author} {\bibfnamefont {H.-V.}\
  \bibnamefont {Nguyen}}, \bibinfo {author} {\bibfnamefont {A.}~\bibnamefont
  {Otero-de-la Roza}}, \bibinfo {author} {\bibfnamefont {L.}~\bibnamefont
  {Paulatto}}, \bibinfo {author} {\bibfnamefont {S.}~\bibnamefont {Poncé}},
  \bibinfo {author} {\bibfnamefont {D.}~\bibnamefont {Rocca}}, \bibinfo
  {author} {\bibfnamefont {R.}~\bibnamefont {Sabatini}}, \bibinfo {author}
  {\bibfnamefont {B.}~\bibnamefont {Santra}}, \bibinfo {author} {\bibfnamefont
  {M.}~\bibnamefont {Schlipf}}, \bibinfo {author} {\bibfnamefont {A.~P.}\
  \bibnamefont {Seitsonen}}, \bibinfo {author} {\bibfnamefont {A.}~\bibnamefont
  {Smogunov}}, \bibinfo {author} {\bibfnamefont {I.}~\bibnamefont {Timrov}},
  \bibinfo {author} {\bibfnamefont {T.}~\bibnamefont {Thonhauser}}, \bibinfo
  {author} {\bibfnamefont {P.}~\bibnamefont {Umari}}, \bibinfo {author}
  {\bibfnamefont {N.}~\bibnamefont {Vast}}, \bibinfo {author} {\bibfnamefont
  {X.}~\bibnamefont {Wu}},\ and\ \bibinfo {author} {\bibfnamefont
  {S.}~\bibnamefont {Baroni}},\ }\bibfield  {title} {\bibinfo {title} {Advanced
  capabilities for materials modelling with {Quantum} {ESPRESSO}},\ }\href
  {https://doi.org/10.1088/1361-648X/aa8f79} {\bibfield  {journal} {\bibinfo
  {journal} {J. Phys. Condens. Matter}\ }\textbf {\bibinfo {volume} {29}},\
  \bibinfo {pages} {465901} (\bibinfo {year} {2017})}\BibitemShut {NoStop}%
\bibitem [{sup()}]{supp}%
  \BibitemOpen
  \href@noop {} {\bibinfo {title} {{See Supplemental Material at [URL will be
  inserted by publisher] for additional results including band structure of
  large-angle tBLG, phonon dispersions in tBLG at different temperatures,
  scaling of the e-ph matrix element with unit cell size, and interlayer versus
  intralayer e-ph coupling in 21.8$^\circ$ tBLG.}}}\BibitemShut {Stop}%
\bibitem [{\citenamefont {Zhou}\ \emph {et~al.}(2021)\citenamefont {Zhou},
  \citenamefont {Park}, \citenamefont {Lu}, \citenamefont {Maliyov},
  \citenamefont {Tong},\ and\ \citenamefont {Bernardi}}]{Zhou2021}%
  \BibitemOpen
  \bibfield  {author} {\bibinfo {author} {\bibfnamefont {J.-J.}\ \bibnamefont
  {Zhou}}, \bibinfo {author} {\bibfnamefont {J.}~\bibnamefont {Park}}, \bibinfo
  {author} {\bibfnamefont {I.-T.}\ \bibnamefont {Lu}}, \bibinfo {author}
  {\bibfnamefont {I.}~\bibnamefont {Maliyov}}, \bibinfo {author} {\bibfnamefont
  {X.}~\bibnamefont {Tong}},\ and\ \bibinfo {author} {\bibfnamefont
  {M.}~\bibnamefont {Bernardi}},\ }\bibfield  {title} {\bibinfo {title}
  {Perturbo: A software package for ab initio electron–phonon interactions,
  charge transport and ultrafast dynamics},\ }\href
  {https://doi.org/https://doi.org/10.1016/j.cpc.2021.107970} {\bibfield
  {journal} {\bibinfo  {journal} {Comput. Phys. Commun.}\ }\textbf {\bibinfo
  {volume} {264}},\ \bibinfo {pages} {107970} (\bibinfo {year}
  {2021})}\BibitemShut {NoStop}%
\bibitem [{\citenamefont {Pizzi}\ \emph {et~al.}(2020)\citenamefont {Pizzi},
  \citenamefont {Vitale}, \citenamefont {Arita}, \citenamefont {Bl{\"{u}}gel},
  \citenamefont {Freimuth}, \citenamefont {G{\'{e}}ranton}, \citenamefont
  {Gibertini}, \citenamefont {Gresch}, \citenamefont {Johnson}, \citenamefont
  {Koretsune}, \citenamefont {Iba{\&}ntilde;ez-Azpiroz}, \citenamefont {Lee},
  \citenamefont {Lihm}, \citenamefont {Marchand}, \citenamefont {Marrazzo},
  \citenamefont {Mokrousov}, \citenamefont {Mustafa}, \citenamefont {Nohara},
  \citenamefont {Nomura}, \citenamefont {Paulatto}, \citenamefont
  {Ponc{\'{e}}}, \citenamefont {Ponweiser}, \citenamefont {Qiao}, \citenamefont
  {Th{\"{o}}le}, \citenamefont {Tsirkin}, \citenamefont {Wierzbowska},
  \citenamefont {Marzari}, \citenamefont {Vanderbilt}, \citenamefont {Souza},
  \citenamefont {Mostofi},\ and\ \citenamefont {Yates}}]{Pizzi2020}%
  \BibitemOpen
  \bibfield  {author} {\bibinfo {author} {\bibfnamefont {G.}~\bibnamefont
  {Pizzi}}, \bibinfo {author} {\bibfnamefont {V.}~\bibnamefont {Vitale}},
  \bibinfo {author} {\bibfnamefont {R.}~\bibnamefont {Arita}}, \bibinfo
  {author} {\bibfnamefont {S.}~\bibnamefont {Bl{\"{u}}gel}}, \bibinfo {author}
  {\bibfnamefont {F.}~\bibnamefont {Freimuth}}, \bibinfo {author}
  {\bibfnamefont {G.}~\bibnamefont {G{\'{e}}ranton}}, \bibinfo {author}
  {\bibfnamefont {M.}~\bibnamefont {Gibertini}}, \bibinfo {author}
  {\bibfnamefont {D.}~\bibnamefont {Gresch}}, \bibinfo {author} {\bibfnamefont
  {C.}~\bibnamefont {Johnson}}, \bibinfo {author} {\bibfnamefont
  {T.}~\bibnamefont {Koretsune}}, \bibinfo {author} {\bibfnamefont
  {J.}~\bibnamefont {Iba{\&}ntilde;ez-Azpiroz}}, \bibinfo {author}
  {\bibfnamefont {H.}~\bibnamefont {Lee}}, \bibinfo {author} {\bibfnamefont
  {J.~M.}\ \bibnamefont {Lihm}}, \bibinfo {author} {\bibfnamefont
  {D.}~\bibnamefont {Marchand}}, \bibinfo {author} {\bibfnamefont
  {A.}~\bibnamefont {Marrazzo}}, \bibinfo {author} {\bibfnamefont
  {Y.}~\bibnamefont {Mokrousov}}, \bibinfo {author} {\bibfnamefont {J.~I.}\
  \bibnamefont {Mustafa}}, \bibinfo {author} {\bibfnamefont {Y.}~\bibnamefont
  {Nohara}}, \bibinfo {author} {\bibfnamefont {Y.}~\bibnamefont {Nomura}},
  \bibinfo {author} {\bibfnamefont {L.}~\bibnamefont {Paulatto}}, \bibinfo
  {author} {\bibfnamefont {S.}~\bibnamefont {Ponc{\'{e}}}}, \bibinfo {author}
  {\bibfnamefont {T.}~\bibnamefont {Ponweiser}}, \bibinfo {author}
  {\bibfnamefont {J.}~\bibnamefont {Qiao}}, \bibinfo {author} {\bibfnamefont
  {F.}~\bibnamefont {Th{\"{o}}le}}, \bibinfo {author} {\bibfnamefont {S.~S.}\
  \bibnamefont {Tsirkin}}, \bibinfo {author} {\bibfnamefont {M.}~\bibnamefont
  {Wierzbowska}}, \bibinfo {author} {\bibfnamefont {N.}~\bibnamefont
  {Marzari}}, \bibinfo {author} {\bibfnamefont {D.}~\bibnamefont {Vanderbilt}},
  \bibinfo {author} {\bibfnamefont {I.}~\bibnamefont {Souza}}, \bibinfo
  {author} {\bibfnamefont {A.~A.}\ \bibnamefont {Mostofi}},\ and\ \bibinfo
  {author} {\bibfnamefont {J.~R.}\ \bibnamefont {Yates}},\ }\bibfield  {title}
  {\bibinfo {title} {{Wannier90 as a community code: new features and
  applications}},\ }\href {https://doi.org/10.1088/1361-648X/AB51FF} {\bibfield
   {journal} {\bibinfo  {journal} {J. Condens. Matter Phys.}\ }\textbf
  {\bibinfo {volume} {32}},\ \bibinfo {pages} {165902} (\bibinfo {year}
  {2020})}\BibitemShut {NoStop}%
\bibitem [{\citenamefont {Pallikara}\ \emph {et~al.}(2022)\citenamefont
  {Pallikara}, \citenamefont {Kayastha}, \citenamefont {Skelton},\ and\
  \citenamefont {Whalley}}]{Pallikara_2022}%
  \BibitemOpen
  \bibfield  {author} {\bibinfo {author} {\bibfnamefont {I.}~\bibnamefont
  {Pallikara}}, \bibinfo {author} {\bibfnamefont {P.}~\bibnamefont {Kayastha}},
  \bibinfo {author} {\bibfnamefont {J.~M.}\ \bibnamefont {Skelton}},\ and\
  \bibinfo {author} {\bibfnamefont {L.~D.}\ \bibnamefont {Whalley}},\
  }\bibfield  {title} {\bibinfo {title} {The physical significance of imaginary
  phonon modes in crystals},\ }\href {https://doi.org/10.1088/2516-1075/ac78b3}
  {\bibfield  {journal} {\bibinfo  {journal} {Electron. Struct.}\ }\textbf
  {\bibinfo {volume} {4}},\ \bibinfo {pages} {033002} (\bibinfo {year}
  {2022})}\BibitemShut {NoStop}%
\bibitem [{\citenamefont {Hellman}\ \emph {et~al.}(2011)\citenamefont
  {Hellman}, \citenamefont {Abrikosov},\ and\ \citenamefont
  {Simak}}]{Hellman2011}%
  \BibitemOpen
  \bibfield  {author} {\bibinfo {author} {\bibfnamefont {O.}~\bibnamefont
  {Hellman}}, \bibinfo {author} {\bibfnamefont {I.~A.}\ \bibnamefont
  {Abrikosov}},\ and\ \bibinfo {author} {\bibfnamefont {S.~I.}\ \bibnamefont
  {Simak}},\ }\bibfield  {title} {\bibinfo {title} {Lattice dynamics of
  anharmonic solids from first principles},\ }\href
  {https://doi.org/10.1103/PhysRevB.84.180301} {\bibfield  {journal} {\bibinfo
  {journal} {Phys. Rev. B}\ }\textbf {\bibinfo {volume} {84}},\ \bibinfo
  {pages} {180301} (\bibinfo {year} {2011})}\BibitemShut {NoStop}%
\bibitem [{\citenamefont {Hellman}\ \emph {et~al.}(2013)\citenamefont
  {Hellman}, \citenamefont {Steneteg}, \citenamefont {Abrikosov},\ and\
  \citenamefont {Simak}}]{Hellman2013}%
  \BibitemOpen
  \bibfield  {author} {\bibinfo {author} {\bibfnamefont {O.}~\bibnamefont
  {Hellman}}, \bibinfo {author} {\bibfnamefont {P.}~\bibnamefont {Steneteg}},
  \bibinfo {author} {\bibfnamefont {I.~A.}\ \bibnamefont {Abrikosov}},\ and\
  \bibinfo {author} {\bibfnamefont {S.~I.}\ \bibnamefont {Simak}},\ }\bibfield
  {title} {\bibinfo {title} {Temperature dependent effective potential method
  for accurate free energy calculations of solids},\ }\href
  {https://doi.org/10.1103/PhysRevB.87.104111} {\bibfield  {journal} {\bibinfo
  {journal} {Phys. Rev. B}\ }\textbf {\bibinfo {volume} {87}},\ \bibinfo
  {pages} {104111} (\bibinfo {year} {2013})}\BibitemShut {NoStop}%
\bibitem [{\citenamefont {Zhou}\ \emph {et~al.}(2018)\citenamefont {Zhou},
  \citenamefont {Hellman},\ and\ \citenamefont {Bernardi}}]{zhou_soft}%
  \BibitemOpen
  \bibfield  {author} {\bibinfo {author} {\bibfnamefont {J.-J.}\ \bibnamefont
  {Zhou}}, \bibinfo {author} {\bibfnamefont {O.}~\bibnamefont {Hellman}},\ and\
  \bibinfo {author} {\bibfnamefont {M.}~\bibnamefont {Bernardi}},\ }\bibfield
  {title} {\bibinfo {title} {Electron-phonon scattering in the presence of soft
  modes and electron mobility in {${\mathrm{SrTiO}}_{3}$} perovskite from first
  principles},\ }\href
  {https://link.aps.org/doi/10.1103/PhysRevLett.121.226603} {\bibfield
  {journal} {\bibinfo  {journal} {Phys. Rev. Lett.}\ }\textbf {\bibinfo
  {volume} {121}},\ \bibinfo {pages} {226603} (\bibinfo {year}
  {2018})}\BibitemShut {NoStop}%
\bibitem [{\citenamefont {Bernardi}(2016)}]{bernardi2016}%
  \BibitemOpen
  \bibfield  {author} {\bibinfo {author} {\bibfnamefont {M.}~\bibnamefont
  {Bernardi}},\ }\bibfield  {title} {\bibinfo {title} {First-principles
  dynamics of electrons and phonons},\ }\href
  {https://link.springer.com/article/10.1140/epjb/e2016-70399-4} {\bibfield
  {journal} {\bibinfo  {journal} {Eur. Phys. J. B}\ }\textbf {\bibinfo {volume}
  {89}},\ \bibinfo {pages} {1} (\bibinfo {year} {2016})}\BibitemShut {NoStop}%
\bibitem [{Note1()}]{Note1}%
  \BibitemOpen
  \bibinfo {note} {We use $10^6$ random $\protect \mathbf {q}$-points sampled
  uniformly in the Brillouin zone and a 2 meV Gaussian smearing for the Dirac
  delta function to ensure the scattering rate is converged down to
  15~K.}\BibitemShut {Stop}%
\bibitem [{\citenamefont {Bernardi}\ \emph {et~al.}(2014)\citenamefont
  {Bernardi}, \citenamefont {Vigil-Fowler}, \citenamefont {Lischner},
  \citenamefont {Neaton},\ and\ \citenamefont {Louie}}]{Marco2014}%
  \BibitemOpen
  \bibfield  {author} {\bibinfo {author} {\bibfnamefont {M.}~\bibnamefont
  {Bernardi}}, \bibinfo {author} {\bibfnamefont {D.}~\bibnamefont
  {Vigil-Fowler}}, \bibinfo {author} {\bibfnamefont {J.}~\bibnamefont
  {Lischner}}, \bibinfo {author} {\bibfnamefont {J.~B.}\ \bibnamefont
  {Neaton}},\ and\ \bibinfo {author} {\bibfnamefont {S.~G.}\ \bibnamefont
  {Louie}},\ }\bibfield  {title} {\bibinfo {title} {Ab initio study of hot
  carriers in the first picosecond after sunlight absorption in silicon},\
  }\href {https://doi.org/10.1103/PhysRevLett.112.257402} {\bibfield  {journal}
  {\bibinfo  {journal} {Phys. Rev. Lett.}\ }\textbf {\bibinfo {volume} {112}},\
  \bibinfo {pages} {257402} (\bibinfo {year} {2014})}\BibitemShut {NoStop}%
\bibitem [{\citenamefont {Stormer}\ \emph {et~al.}(1990)\citenamefont
  {Stormer}, \citenamefont {Pfeiffer}, \citenamefont {Baldwin},\ and\
  \citenamefont {West}}]{Stormer1990}%
  \BibitemOpen
  \bibfield  {author} {\bibinfo {author} {\bibfnamefont {H.~L.}\ \bibnamefont
  {Stormer}}, \bibinfo {author} {\bibfnamefont {L.~N.}\ \bibnamefont
  {Pfeiffer}}, \bibinfo {author} {\bibfnamefont {K.~W.}\ \bibnamefont
  {Baldwin}},\ and\ \bibinfo {author} {\bibfnamefont {K.~W.}\ \bibnamefont
  {West}},\ }\bibfield  {title} {\bibinfo {title} {Observation of a
  {Bloch-Gr\"uneisen} regime in two-dimensional electron transport},\ }\href
  {https://doi.org/10.1103/PhysRevB.41.1278} {\bibfield  {journal} {\bibinfo
  {journal} {Phys. Rev. B}\ }\textbf {\bibinfo {volume} {41}},\ \bibinfo
  {pages} {1278} (\bibinfo {year} {1990})}\BibitemShut {NoStop}%
\bibitem [{Note2()}]{Note2}%
  \BibitemOpen
  \bibinfo {note} {At temperatures below $T_\protect \mathrm {BG}$, which are
  not consider here, the resistivity approaches a $\rho \propto T^4$ behavior
  in the low-temperature limit~\cite {Hwang2008}.}\BibitemShut {Stop}%
\bibitem [{\citenamefont {Dean}\ \emph {et~al.}(2010)\citenamefont {Dean},
  \citenamefont {Young}, \citenamefont {Meric}, \citenamefont {Lee},
  \citenamefont {Wang}, \citenamefont {Sorgenfrei}, \citenamefont {Watanabe},
  \citenamefont {Taniguchi}, \citenamefont {Kim}, \citenamefont {Shepard},\
  and\ \citenamefont {Hone}}]{Dean2010}%
  \BibitemOpen
  \bibfield  {author} {\bibinfo {author} {\bibfnamefont {C.~R.}\ \bibnamefont
  {Dean}}, \bibinfo {author} {\bibfnamefont {A.~F.}\ \bibnamefont {Young}},
  \bibinfo {author} {\bibfnamefont {I.}~\bibnamefont {Meric}}, \bibinfo
  {author} {\bibfnamefont {C.}~\bibnamefont {Lee}}, \bibinfo {author}
  {\bibfnamefont {L.}~\bibnamefont {Wang}}, \bibinfo {author} {\bibfnamefont
  {S.}~\bibnamefont {Sorgenfrei}}, \bibinfo {author} {\bibfnamefont
  {K.}~\bibnamefont {Watanabe}}, \bibinfo {author} {\bibfnamefont
  {T.}~\bibnamefont {Taniguchi}}, \bibinfo {author} {\bibfnamefont
  {P.}~\bibnamefont {Kim}}, \bibinfo {author} {\bibfnamefont {K.~L.}\
  \bibnamefont {Shepard}},\ and\ \bibinfo {author} {\bibfnamefont
  {J.}~\bibnamefont {Hone}},\ }\bibfield  {title} {\bibinfo {title} {Boron
  nitride substrates for high-quality graphene electronics},\ }\href
  {https://www.nature.com/articles/nnano.2010.172} {\bibfield  {journal}
  {\bibinfo  {journal} {Nat. Nanotechnol.}\ }\textbf {\bibinfo {volume} {5}},\
  \bibinfo {pages} {722} (\bibinfo {year} {2010})}\BibitemShut {NoStop}%
\bibitem [{\citenamefont {Efetov}\ and\ \citenamefont
  {Kim}(2010)}]{Efetov2010}%
  \BibitemOpen
  \bibfield  {author} {\bibinfo {author} {\bibfnamefont {D.~K.}\ \bibnamefont
  {Efetov}}\ and\ \bibinfo {author} {\bibfnamefont {P.}~\bibnamefont {Kim}},\
  }\bibfield  {title} {\bibinfo {title} {Controlling electron-phonon
  interactions in graphene at ultrahigh carrier densities},\ }\href
  {https://link.aps.org/doi/10.1103/PhysRevLett.105.256805} {\bibfield
  {journal} {\bibinfo  {journal} {Phys. Rev. Lett.}\ }\textbf {\bibinfo
  {volume} {105}},\ \bibinfo {pages} {256805} (\bibinfo {year}
  {2010})}\BibitemShut {NoStop}%
\bibitem [{Note3()}]{Note3}%
  \BibitemOpen
  \bibinfo {note} {Of the three degenerate modes at K, the mode shown in Fig.
  4(c) is the one with a downward dispersion along the K$-$$\Gamma $ and M$-$K
  directions. This is the mode with the strongest coupling near K.}\BibitemShut
  {Stop}%
\end{thebibliography}
\end{document}